\documentclass[final,3p,times]{elsarticle}

\usepackage{natbib}           	 																				
\usepackage{graphicx}          																				
\usepackage{amsmath,amsfonts,amssymb}														%
\usepackage{algorithmic}																					%
\usepackage{algorithm}																						%
\usepackage{amstext}																							%
\usepackage{psfrag}																							%
\usepackage{float}																								%
\usepackage{booktabs}																						%
\usepackage[table]{xcolor}																					%
\usepackage{multirow}																						%
\usepackage{multicol}																							%
\usepackage{stackrel}																							%
\usepackage{stfloats}																							%
\usepackage{multirow}																						%
\usepackage{arydshln}																							%
\usepackage{color}																								%
\usepackage{tikz}																								%
\usepackage{pgfplots}																							%
\usepackage{url}																									%
\usepackage{pgf}																									%
\usetikzlibrary{arrows,automata,matrix,fit,positioning,calc}								%
\usepackage[latin1]{inputenc}																				%
\usepackage{verbatim}																						%

\newtheorem{thm}{Theorem}[section]

\newtheorem{prop}[thm]{Proposition}
\newtheorem{defi}[thm]{Definition}
\newtheorem{remark}{Remark}

\journal{Nonlinear Analysis: Hybrid Systems}

\begin{document}

\begin{frontmatter}

\title{Deterministic and Stochastic Approaches to Supervisory Control Design for Networked Systems with Time-Varying Communication Delays}

\author[KTH]{Burak Demirel\corref{cor1}}
\ead{demirel@ee.kth.se}
\ead[url]{www.ee.kth.se/~demirel}
\cortext[cor1]{Corresponding author}
\author[ETH]{Corentin Briat}
\ead{briatc@bsse.ethz.ch}
\ead[url]{www.briat.info}
\author[KTH]{Mikael Johansson}
\ead{mikaelj@ee.kth.se}
\ead[url]{www.s3.kth.se/~mikaelj}

\address[KTH]{Automatic Control Laboratory, School of Electrical Engineering, KTH Royal Institute of Technology}
\address[ETH]{Control Theory and Systems Biology, Department of Biosystems Science and Engineering, Swiss Federal Institute of Technology Z\"{u}rich}

\begin{abstract}
This paper proposes a supervisory control structure for networked systems with time-varying delays. The control structure, in which a supervisor triggers the most appropriate controller from a multi-controller unit, aims at improving the closed-loop performance relative to what can be obtained using a single robust controller. Our analysis considers average dwell-time switching and is based on a novel multiple Lyapunov-Krasovskii functional. We develop stability conditions that can be verified by semi-definite programming, and show that the associated state feedback synthesis problem also can be solved using convex optimization tools. Extensions of the analysis and synthesis procedures to the case when the evolution of the delay mode is described by a Markov chain are also developed. Simulations on small and large-scale networked control systems are used to illustrate the effectiveness of our approach.
\end{abstract}


\end{frontmatter}

\section{Introduction}\label{sec:intro}

Networked control systems are distributed systems that use communication networks to exchange information between sensors, controllers and actuators~\cite{ZBP:01,WYB:02}. The networked control system architecture promises advantages in terms of increased flexibility, reduced wiring and lower maintenance costs, and is finding its way into a wide range of applications, from automobiles and transportation to  process control and power systems, see \emph{e.g.},~\cite{WYB:02}~--\nocite{MoT:07, HNX:07}~\cite{NSS+:05}. 

\begin{figure}\centering
  	\includegraphics{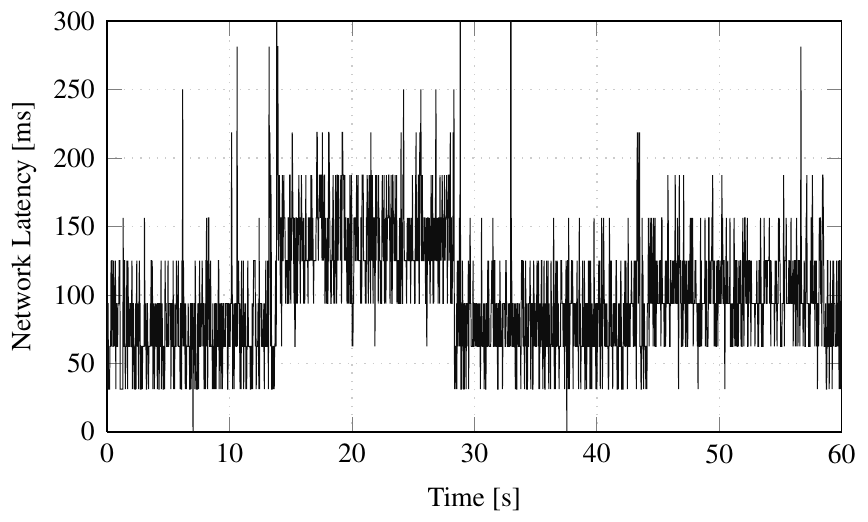}
    \caption{The figure shows a recorded delay trace from the multi-hop wireless networking protocol used for networked control in~\cite{WPJ+:07}. The delay exhibits distinct mode changes (here corresponding to one, two or three-hop communication) and varies around its piecewise constant mode-dependent mean. Similar behavior was reported by~\cite{KTK+:08}, who measured the delay of sensor data sent over a CAN bus. Their delay varied between 10-20~ms, but increased abruptly to around 150~ms under certain network conditions.}
    \label{fig:experimental_delay}
\end{figure}

The use of a shared communication medium introduces time-varying delays and information losses which may deteriorate the system's performance, even to the point where the closed-loop system becomes unstable.  A conservative approach is to design a robust controller that considers the worst-case delay. However, this might cause poor performance if the actual delay is only rarely close to its upper bound. Therefore, there is currently a renewed interest in adapting the control law to the delay evolution (\emph{e.g.},~\cite{CHB:06}~--\nocite{HCB:06, JFK+:09,DBJ:12,HDR+:11}~\cite{KJF+:12}). Inspired by the communication delays that we have experienced in applications, see Figure~\ref{fig:experimental_delay}, we design a supervisory control scheme in the sense of~\cite{Mor:96}. This control architecture consists of a finite number of controllers, each designed for a bounded delay variation (corresponding, \emph{e.g.}, to low, medium and high network load) and a supervisor which orchestrates the switching among them.

The analysis of switched systems with fixed time-delays is challenging and has attracted significant attention in the literature, \emph{e.g.},~\cite{HCB:06,XiW:05,SZH:06,YaO:08}. Only recently, however, attempts to analyze switched systems with \emph{time-varying} delays have begun to appear. Distinctively,~\cite{JFK+:09} constructed multiple Lyapunov-Krasovskii  functionals that guarantee closed-loop stability under a minimum dwell-time condition for interval time-varying delays. An alternative approach to deal with time-varying delays is to assume that they evolve according to a Markov chain and develop conditions that ensure (mean-square) stability, see~\emph{e.g.},~\cite{Nil:98}~--\nocite{XHH:00,YWH+:06,BLL:10,WCW:06}~\cite{BeB:98}. The work in~\cite{Nil:98} assumed that the time delay never exceeds a sampling interval, modeled its evolution as a Markov process,  and derived the associated LQG-optimal controller. However, this formulation is not able to deal with longer time-delays. The work in~\cite{XHH:00} proposed a discrete-time Markovian jump linear system formulation, which allows longer (but bounded) time delays, and posed the design of a mode-dependent controller as a non-convex optimization problem. Complementary to these discrete-time formulations,~\cite{BLL:10}~--\nocite{WCW:06}~\cite{BeB:98} have investigated the mean-square stability of continuous-time linear systems with random time delays using stochastic Lyapunov-Krasovskii functionals. The papers~\cite{CHB:08, CHB:08b} have applied these techniques to networked control systems with random communication delays and synthesized mode-dependent controllers.

In this paper, we analyze our proposed supervisory control structure by combining a novel multiple Lyapunov-Krasovskii functional with the assumption of average dwell-time switching. The average dwell-time concept, introduced in \cite{HeM:99}, is a natural deterministic abstraction of load changes in communication networks, where minimal or maximal durations for certain traffic conditions are hard to guarantee. We demonstrate that the existence of a multiple Lyapunov-Krasovskii functional that ensures closed-loop stability under average dwell-time switching can be verified by solving a set of linear matrix inequalities. In addition, we show that the state feedback synthesis problem for the proposed supervisory control structure can also be solved via semi-definite programming. A similar analysis for Markovian time-delays is developed based on a slightly less powerful Lyapunov-Krasovskii functional than the one underpinning our deterministic analysis. Also in this case, we manage to design mode-dependent state feedback controllers using convex optimization.

The organization of the paper is as follows. Section~2.1 presents the supervisory control structure and formalizes the relevant analysis and synthesis problems in a deterministic setting. In Section 2.2, multiple Lyapunov-Krasovskii functionals are constructed for establishing exponential stability of supervisory control systems under average dwell-time switchings. Additionally, LMI conditions that verify the existence of such a multiple Lyapunov functional are derived. State-feedback synthesis conditions are also given in Section~2.3.  Section~3 develops a similar analysis framework for stochastic delays. Section~3.1 formulates switched control system problem introduced in Section~2 as a Markovian jump linear system. Section~3.2 develops stochastic exponential mean-square stability conditions for the supervisory control system under stochastic delays. The corresponding state-feedback synthesis conditions are proposed in Section 3.3. Numerical examples are used to demonstrate the effectiveness of the proposed techniques in Section~4. Finally, Section~5 concludes the paper.

\emph{Notation:} Throughout this paper, $\mathbb{R}^{n}$ denotes the \emph{n}--dimensional Euclidean space, $\mathbb{R}^{m \times n}$ is the set of all $m \times n$ real matrices, and $\mathbb{S}^{n}_{++}$ denotes the cone of real symmetric positive definite matrices of dimension $n$. For a real square matrix $M$ we define $M^{\mathcal{S}}\triangleq M+M^{\intercal}$ where $M^{\intercal}$ is its transpose. Additionally, '$\star$' represents symmetric terms in symmetric matrices and in quadratic forms, $\otimes$ denotes the Kronecker product, and $\mathbb{R}_{\geq 0}~\big(\mathbb{R}_{>0}\big)$ is the set of nonnegative (positive) real numbers. Lastly, $\mathrm{col}(\lambda_{i})$ is the column vector with components $\lambda_{i}$. 

\begin{figure}\centering
    \includegraphics[angle=0,width=0.5\columnwidth]{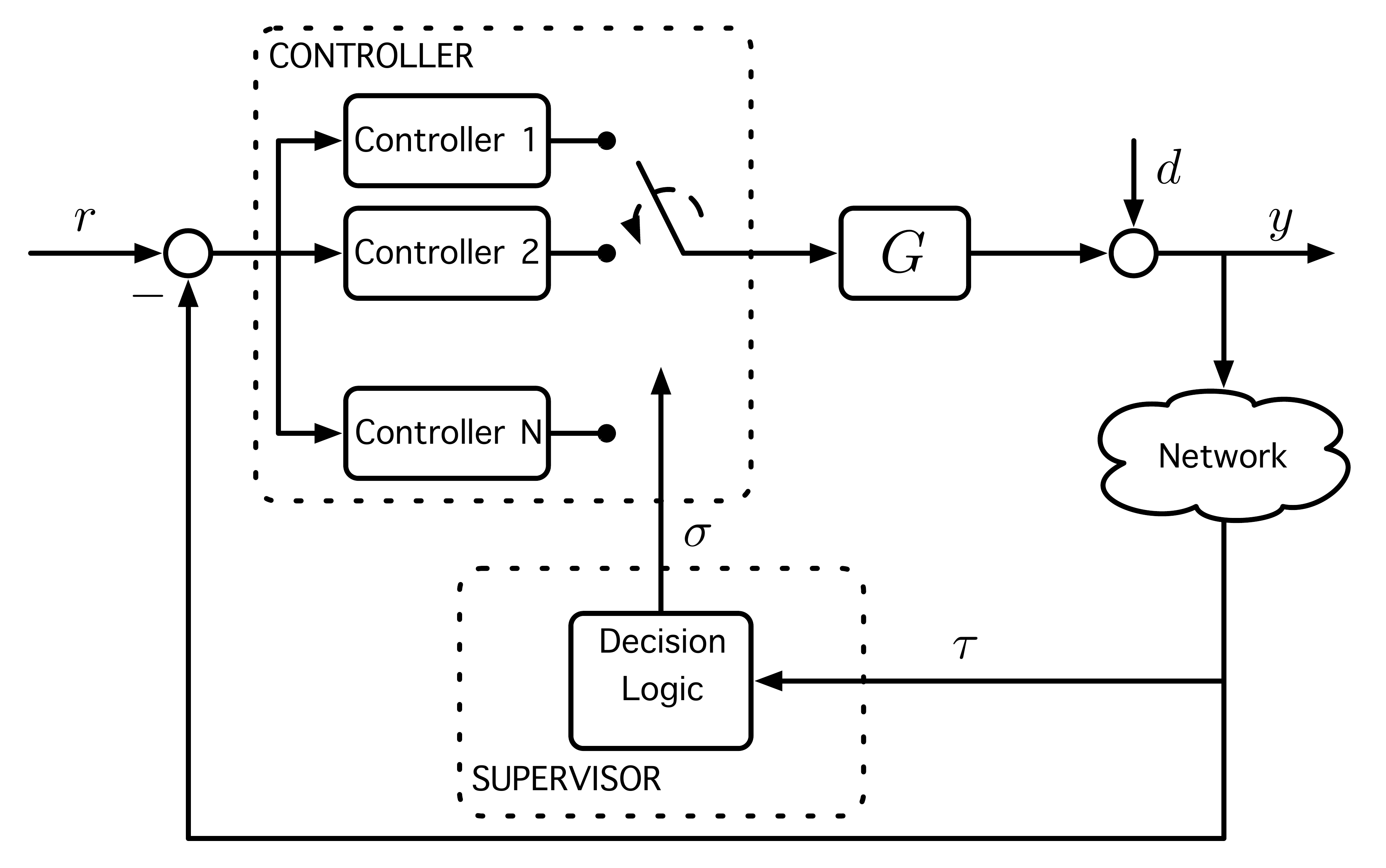}\\
    \caption{The general block scheme of the proposed supervisory control structure.}
    \label{fig:control_structure}
\end{figure}

\section{Deterministic Switched Systems}

\subsection{System Modeling}\label{sec:sysmodel}

We consider the supervisory control system in Figure 2. Here, $G$ is the plant to be controlled, described by 
\begin{equation}
\dot{x}(t) = Ax(t) + Bu(t)
\end{equation}
where $A\in \mathbb{R}^{n\times n}$ and $B\in \mathbb{R}^{n\times m}$. The network is modelled as a time-varying delay $\tau_{\sigma(t)}(t)$ where $\sigma:\mathbb{R}_{\geq0}\mapsto\mathcal{M}$ with $\mathcal{M}=\{1,\ldots,M\}$ is the mode (operating condition) of the network. We assume that the delay in each mode is bounded, 
\begin{equation*}
h_{1}\leq\underline{h}_{\sigma(t)}\leq\tau_{\sigma(t)}(t)\leq \overline{h}_{\sigma(t)}\leq h_{M+1}\;.
\end{equation*}
The multi-controller unit uses the mode signal to select and apply the corresponding mode-dependent feedback law
\begin{equation}
u(t) = K_{\sigma(t)}x\big( t-\tau_{\sigma(t)}(t) \big).
\end{equation} 
In this way, the closed-loop system is described by the following switched linear system with time-varying delay
\begin{align}
\setlength{\arraycolsep}{2.0pt}
\begin{array}{rll} 
\Sigma_{1}: & \dot{x}(t)=Ax(t)+A_{\sigma(t)}x\big(t-\tau_{\sigma(t)}(t)\big), & \forall t\in\mathbb{R}_{\geq0} \\ \vspace*{1mm}
 & x(t) = \varphi(t), & \forall t\in [-h_{M+1},0]
\end{array}\label{eq:SwitchedSys}
\end{align}
where $A_{\sigma(t)}\triangleq BK_{\sigma(t)}\in\mathbb{R}^{n\times n}$ and  $\varphi(t)\in \mathcal{C}\big([-h_{M+1},0],\mathbb{R}^{n}\big)$ is the initial function belonging to $\mathcal{C}\big([-h_{M+1},0],\mathbb{R}^{n}\big)$, the Banach space of continuous functions defined on $[-h_{M+1},0]$.

\begin{defi}
The system~\eqref{eq:SwitchedSys} is exponentially stable under the switching signal $\sigma (t)$ if there exist positive constants $\gamma$ and $\alpha$ such that the solution of $x(t)$ of the system~\eqref{eq:SwitchedSys} satisfies
\begin{equation*}
\parallel x(t)\parallel \; \leq\gamma\parallel x(t_{0})\parallel_{\mathcal{C}}~e^{-\alpha (t-t_{0})}, \quad t\geq t_{0}
\end{equation*}
where $\parallel x(t_{0})\parallel_{\mathcal{C}}\triangleq \underset{-h_{M+1}\leq\theta\leq 0}{\sup}\{\parallel x(t_{0}+\theta)\parallel,\parallel\dot{x}(t_{0}+\theta)\parallel\}$.
\end{defi}

In order to guarantee exponential stability, we will put restrictions on the switching signal $\sigma(t)$. Specifically, we will assume that the signal satisfies an average dwell-time condition in the following sense.

\begin{defi}[Liberzon~\cite{Lib:03}]
We denote the number of jumps of a switching signal~$\sigma$ on the interval $(t,T)$ by $N_{\sigma}(T,t)$. Then we say that~$\sigma$ has the \emph{average dwell-time} $\tau_{a}$ if there exist two positive numbers $N_{0}$ and $\tau_{a}$ such that
\begin{align*}
N_{\sigma}(T,t)\leq N_{0}+\frac{T-t}{\tau_{a}}\,,\quad \forall T>t\geq 0\,.
\end{align*}
The set of all switching signals satisfying the above condition is denoted by $\mathcal{S}[\tau_{a}]$.
\end{defi}
We consider two specific problems in this section. The first is to verify that the switched linear system~\eqref{eq:SwitchedSys} is exponentially stable under average dwell-time switching. The second one is to design state feedback controllers for each mode such that the supervisory control system is exponentially stable with guaranteed convergence rate.

\subsection{Exponential Stability Analysis Using Multiple Lyapunov -- Krasovskii Functionals}\label{subsec:DeterStability}
The exponential stability of switched system~\eqref{eq:SwitchedSys} is equivalent to the existence of a scalar $\alpha\in\mathbb{R}_{\geq0}$ such that $e^{\alpha t}||x(t)||$ converges asymptotically to zero for each $\sigma\in\mathcal{S}[\tau_{a}]$. To characterize the rate of convergence of the system~\eqref{eq:SwitchedSys}, let us consider the change of variables $\xi(t)\triangleq e^{\alpha t}x(t)$. Then we have:
\begin{align}
\dot{\xi}(t) =\alpha e^{\alpha t}x(t)+e^{\alpha t}\dot{x}(t) & = \alpha\xi(t)+e^{\alpha t}\big[ Ax(t)+A_{i}x\big(t-\tau_{i}(t)\big) \big] \nonumber\\
                 &=(\alpha I_{n}+A)\xi(t)+e^{\alpha \tau_{i}(t)}A_{i}\xi\big(t-\tau_{i}(t)\big)\,, \label{SwitchedSysTV}
\end{align}
where $\tau_{i}(t)\in[h_{i},h_{i+1})~\forall i\in\mathcal{M}$. Note that the asymptotic stability of~\eqref{SwitchedSysTV} implies that the original system is exponentially stable with decay rate $\alpha$.
However, this change of variables introduces a time-varying coefficient in the switched system model~\eqref{SwitchedSysTV}. Similar to~\cite{SDR:04}, we can exploit the (mode-dependent) bounds on the delay and rewrite~\eqref{SwitchedSysTV} in a polytopic form. Specifically, we express the term $e^{\alpha\tau_{i}(t)}$ as a convex combination of the bounds $e^{\alpha h_{i}}$ and $e^{\alpha h_{i+1}}$:
\begin{equation*}
e^{\alpha\tau_{i}(t)}=\lambda_{1}(t)e^{\alpha h_{i}}+\lambda_{2}(t)e^{\alpha h_{i+1}}\,,\quad \forall i\in\mathcal{M} \;,
\end{equation*}
where $\lambda_{1}(t),\lambda_{2}(t)\in\mathbb{R}_{\geq 0}$ and $\lambda_{1}(t)+\lambda_{2}(t)=1,~\forall t\in\mathbb{R}_{\geq 0}$. The delayed differential equation~\eqref{eq:SwitchedSys} is then rewritten as
\begin{align}
\dot{\xi}(t)=A_{\alpha}\xi(t)+\sum_{j=1}^{2}\lambda_{j}(t)A_{\alpha_{ij}}\xi\big(t-\tau_{i}(t)\big) \;, \label{eq:SwitchedSysTV2}
\end{align}
where $A_{\alpha}\triangleq (\alpha I_{n}+A)$ and $A_{\alpha_{ij}}\triangleq\varrho_{ij}A_{i}$~with~$\varrho_{ij}\triangleq e^{\alpha h_{i+j-1}}$ when $\tau_{i}(t)\in [h_{i},h_{i+1}),~\forall i\in\mathcal{M}$.

We combine a novel multiple Lyapunov-Krasovskii functional with the dwell-time approach of~\cite{HeM:99} to establish exponential stability of the switched system~\eqref{eq:SwitchedSys}. For ease of notation, we state the theorem for the case of two delay modes only, but the approach extends immediately to a system with $M$ modes.

\begin{thm}\label{thm:Switched_System_Stability}
There exists a finite constant $\tau_{a}$ such that the switched linear system~\eqref{eq:SwitchedSys} is exponentially stable over $\mathcal{S}[\tau_{a}]$ with a given decay rate $\alpha>0$ for time-varying delays $\tau_{i}(t)\in[h_{i},h_{i+1})$, $\forall i\in\{1,2\}$ if there exist real matrices $P_{i}, Q_{ik}, R_{ik}, S_{ik}, T_{ik}\in\mathbb{S}_{++}^{n}$ and $Z_{ik}\in\mathbb{R}^{n\times n},~\forall i,k\in\{1,2\}$ and a constant scalar $\mu>1$ satisfying $P_{i}\leq\mu P_{j}$, $Q_{ik}\leq\mu Q_{jk}$, $R_{ik}\leq\mu R_{jk}$, $S_{ik}\leq\mu S_{jk}$ and $T_{ik}\leq\mu T_{jk}$, $\forall i, j,k\in\{1,2\}$ such that the LMIs 
\begin{align}
\renewcommand{\arraystretch}{1.2}
\left[\begin{array}{c:cc:ccc:ccccccc}
\Phi_{1} & P_{1}A_{\alpha_{1j}} & 0 & S_{11} & S_{12} & 0 & h_{1}A_{\alpha}^{\intercal}S_{11} & \delta_{1}A_{\alpha}^{\intercal}T_{11} & h_{2}A_{\alpha}^{\intercal}S_{12} & \delta_{2}A_{\alpha}^{\intercal}T_{12}\\ \hdashline
\star & -\Upsilon_{11}^{\mathcal{S}} & 0 & \Upsilon_{11}^{\intercal} & \Upsilon_{11} & 0 & h_{1}A_{\alpha_{1j}}^{\intercal}S_{11} & \delta_{1}A_{\alpha_{1j}}^{\intercal}T_{11} & h_{2}A_{\alpha_{1j}}^{\intercal}S_{12} & \delta_{2}A_{\alpha_{1j}}^{\intercal}T_{12} \\
\star & \star & -\Upsilon_{12}^{\mathcal{S}} & 0 & \Upsilon_{12}^{\intercal} & \Upsilon_{12} & 0 & 0 & 0 & 0 \\ \hdashline
\star & \star & \star & -\Xi_{11} & Z_{11} & 0 & 0 & 0 & 0 & 0 \\
\star & \star & \star & \star & -\Xi_{12} & Z_{12} & 0 & 0 & 0 & 0\\
\star & \star & \star & \star & \star & -\Xi_{13}  & 0 & 0 & 0 & 0 \\ \hdashline 
\star & \star & \star & \star & \star & \star & -S_{11} & 0 & 0 & 0 \\
\star & \star & \star & \star & \star & \star & \star & -T_{11} & 0 & 0 \\
\star & \star & \star & \star & \star & \star & \star & \star & -S_{12} & 0 \\
\star & \star & \star & \star & \star & \star & \star & \star & \star & -T_{12}
\end{array}\right] < 0\label{eq:thm1a} \\
\renewcommand{\arraystretch}{1.2}
\left[\begin{array}{c:cc:ccc:ccccccc}
\Phi_{2} & 0 &  P_{2}A_{\alpha_{2j}} & S_{21} & S_{22} & 0 & h_{1}A_{\alpha}^{\intercal}S_{21} & \delta_{1}A_{\alpha}^{\intercal}T_{21} & h_{2}A_{\alpha}^{\intercal}S_{22} & \delta_{2}A_{\alpha}^{\intercal}T_{22}\\ \hdashline
\star & -\Upsilon_{21}^{\mathcal{S}} & 0 & \Upsilon_{21}^{\intercal} & \Upsilon_{21} & 0  & 0 & 0 & 0 & 0 \\
\star & \star & -\Upsilon_{22}^{\mathcal{S}} & 0 & \Upsilon_{22}^{\intercal} &\Upsilon_{22} & h_{1}A_{\alpha_{2j}}^{\intercal}S_{21} & \delta_{1}A_{\alpha_{2j}}^{\intercal}T_{21} & h_{2}A_{\alpha_{2j}}^{\intercal}S_{22} & \delta_{2}A_{\alpha_{2j}}^{\intercal}T_{22} \\ \hdashline
\star & \star & \star &  -\Xi_{21} & Z_{21} & 0 & 0 & 0 & 0 & 0 \\
\star & \star & \star & \star & -\Xi_{22} & Z_{22} & 0 & 0 & 0 & 0\\
\star & \star & \star & \star & \star & -\Xi_{23} & 0 & 0 & 0 & 0 \\ \hdashline
\star & \star & \star & \star & \star & \star & -S_{21} & 0 & 0 & 0 \\
\star & \star & \star & \star & \star & \star & \star & -T_{21} & 0 & 0 \\
\star & \star & \star & \star & \star & \star & \star & \star & -S_{22} & 0 \\
\star & \star & \star & \star & \star & \star & \star & \star & \star & -T_{22}
\end{array}\right] < 0\label{eq:thm1b} 
\end{align}
\begin{align}
\begin{bmatrix}
T_{ij}&Z_{ij} \\ \star&T_{ij}
\end{bmatrix} \geq 0 \label{eq:thm1c} 
\end{align}
where $\Phi_{i}=P_{i}A_{\alpha}+A_{\alpha}^{\intercal}P_{i}+\sum_{k=1}^{2}\big(Q_{ik}+R_{ik}-S_{ik}\big)$,~$\Upsilon_{i1}=T_{i1}-Z_{i1}$,
~$\Upsilon_{i2}=T_{i2}-Z_{i2}$,~$\Xi_{i1}=Q_{i1}+T_{i1}+S_{i1}$,~$\Xi_{i2}=Q_{i2}+S_{i2}+\sum_{k=1}^{2}T_{ik}+R_{i1}$~and
~$\Xi_{i3}=R_{i2}+T_{i2}$ hold for all $i,j\in\{1,2\}$.
\end{thm}

\noindent\textbf{Proof.} Our claim follows if we can find Lyapunov-Krasovskii functionals $V_{i}(t)$ that guarantee decay rate $\alpha$ while in mode $i$ and a constant $\mu >1$ such that $V_{i}(t)\leq\mu V_{j}(t)~\forall i,j\in\{1,2\}$. Then, by~\cite[Theorem~1]{HeM:99},~\eqref{eq:SwitchedSys} is exponentially stable for every switching signal $\sigma$ with average dwell-time $\tau_{a}>\tau_{a}^{\ast}=\frac{\ln\mu}{\alpha}$.

We consider the following Lyapunov-Krasovskii functional, inspired from~\cite{Sha:08}, $V_{i}:\mathbb{R}^{n}\rightarrow\mathbb{R}_{\geq 0},~\forall i\in\mathcal{M}$ :
\begin{multline}
V_{i}(t)=\xi^{\intercal}(t)P_{i}\xi(t) + \sum_{k=1}^{2}\int_{t-h_{k}}^{t}\xi^{\intercal}(s)Q_{ik}\xi(s)ds +\sum_{k=1}^{2}\int_{t-h_{k+1}}^{t}\xi^{\intercal}(s)R_{ik}\xi(s)ds \\
+ \sum_{k=1}^{2}\int_{-h_{k}}^{0}\int_{t+s}^{t}h_{k}\dot{\xi}^{\intercal}(\theta)S_{ik}\dot{\xi}(\theta)d\theta ds 
+ \sum_{k=1}^{2}\int_{-h_{k+1}}^{-h_{k}}\int_{t+s}^{t}\overbrace{(h_{k+1}-h_{k})}^{\displaystyle{\delta_{k}}}\dot{\xi}^{\intercal}(\theta)T_{ik}\dot{\xi}(\theta)d\theta ds \;. \label{eq:LyapCand}
\end{multline}
The derivative of $V_{i}(t)$ along the trajectory of the system is given by
\begin{multline}
\dot{V}_{i}(t) = 2\dot{\xi}^{\intercal}(t)P_{i}\xi(t) + \xi^{\intercal}(t)\bigg[\sum_{k=1}^{2}\big( Q_{ik}+R_{ik} \big)\bigg]\xi(t) -\sum_{k=1}^{2}\xi^{\intercal}(t-h_{k})Q_{ik} \xi(t-h_{k})
-\sum_{k=1}^{2}\xi^{\intercal}(t-h_{k+1})R_{ik}\xi(t-h_{k+1}) \\ 
+\xi^{\intercal}(t)\bigg[\sum_{j=1}^{2}\big(h_{k}^{2}S_{ik}+\delta_{k}^{2}T_{ik}\big)\bigg]\xi(t) - \sum_{k=1}^{2}\int_{t-h_{k}}^{t}h_{k}\dot{\xi}^{\intercal}(s)S_{ik}\dot{\xi}(s)ds - \sum_{k=1}^{2}\int_{t-h_{k+1}}^{t-h_{k}}\delta_{k}\dot{\xi}^{\intercal}(s)T_{ik}\dot{\xi}(s)ds \;. \label{eq:dotV}
\end{multline}
Using Jensen's inequality~\cite{GKC:03}, the integral term $\int_{t-h_{k}}^{t}h_{k}\dot{\xi}^{\intercal}(s)S_{ik}\dot{\xi}(s)ds$ in the preceding equality is bounded as
\begin{equation}
-\int_{t-h_{k}}^{t}h_{k}\dot{\xi}^{\intercal}(s) S_{ik}\dot{\xi}(s)ds \leq -\big[\xi(t)-\xi(t-h_{k})\big]^{\intercal}S_{ik}\big[\xi(t)-\xi(t-h_{k})\big] \;.
\label{eq:Jensen}
\end{equation}
To upperbound the integral term $\int_{t-h_{k+1}}^{t-h_{k}}\delta_{k}\dot{\xi}^{\intercal}(s)T_{ik}\dot{\xi}(s)ds$, we use the reciprocally convex combination from~\cite{PKJ:11}:
\begin{align}
-\int_{t-h_{k+1}}^{t-h_{k}}\delta_{k}\dot{\xi}^{\intercal}(s) T_{ik}\dot{\xi}(s)ds = &\; -\int_{t-h_{k+1}}^{t-\tau_{k}(t)}\delta_{k}\dot{\xi}^{\intercal}(s)T_{ik}\dot{\xi}(s)ds 
-\int_{t-\tau_{k}(t)}^{t-h_{k}}\delta_{k}\dot{\xi}^{\intercal}(s)T_{ik}\dot{\xi}(s)ds \;, \nonumber \\
\leq &\; -
\begin{bmatrix} \xi(t-h_{k})-\xi(t-\tau_{k}(t)) \\ \xi(t-\tau_{k}(t))-\xi(t-h_{k+1}) \end{bmatrix}^{\intercal} 
\underbrace{\begin{bmatrix} T_{ik} & Z_{ik} \\ \star & T_{ik} \end{bmatrix}}_{\geq 0}
\begin{bmatrix} \xi(t-h_{k})-\xi(t-\tau_{k}(t)) \\ \xi(t-\tau_{k}(t))-\xi(t-h_{k+1}) \end{bmatrix}\;.
\label{eq:Reciprocal}
\end{align}
Substituting~\eqref{eq:Jensen} and~\eqref{eq:Reciprocal} into~\eqref{eq:dotV}, we compute an upper bound for the Lyapunov functional~\eqref{eq:LyapCand} as
\begin{align*}
\renewcommand{\arraystretch}{2}
\setlength{\arraycolsep}{1.2pt}
\begin{array}{rl}
\dot{V}_{i}(t)\leq  & \xi^{\intercal}(t)\bigg[A_{\alpha}^{\intercal}P_{i}+P_{i}A_{\alpha}+\sum_{k=1}^{2}\big(Q_{ik}+R_{ik}-S_{ik}\big) + A_{\alpha}^{\intercal}\sum_{k=1}^{2}\big(h_{k}^{2}S_{ik}+\delta_{k}^{2}T_{ik}\big)A_{\alpha}\bigg]\xi(t) \\
 & +2\xi^{\intercal}(t)\bigg[P_{i}\sum_{j=1}^{2}\big(\lambda_{j}(t)A_{\alpha_{ij}}\big) + A_{\alpha}^{\intercal}\sum_{k=1}^{2}\big(h_{k}^{2}S_{ik} + \delta_{k}^{2}T_{ik}\big)\sum_{j=1}^{2}\big(\lambda_{j}(t)A_{\alpha_{ij}}\big)\bigg]\xi(t-\tau_{i}(t)) \\
 & +\xi^{\intercal}(t-\tau_{i}(t))\sum_{j=1}^{2}\big(\lambda_{j}(t)A_{\alpha_{ij}}\big)^{\intercal}\bigg[\sum_{k=1}^{2}\big(h_{k}^{2}S_{ik}+\delta_{k}^{2}T_{ik}\big)\bigg]\sum_{j=1}^{2}\big(\lambda_{j}(t)A_{\alpha_{ij}}\big)\xi(t-\tau_{i}(t)) \\
  & -\sum_{k=1}^{2}\xi^{\intercal}(t-\tau_{k}(t))\Big(2T_{ik}-Z_{ik}-Z_{ik}^{\intercal}\Big)\xi(t-\tau_{k}(t)) +2\xi^{\intercal}(t)\sum_{k=1}^{2}S_{ik}\xi(t-h_{k}) \\
  & -\sum_{k=1}^{2}\xi^{\intercal}(t-h_{k})\Big(Q_{ik}+S_{ik}+T_{ik}\Big)\xi(t-h_{k}) - \sum_{k=1}^{2}\xi^{\intercal}(t-h_{k+1})\Big(R_{ik}+T_{ik}\Big)\xi(t-h_{k+1}) \\
  & +2\sum_{k=1}^{2}\xi^{\intercal}(t-\tau_{k}(t))\Big(T_{ik}-Z_{ik}\Big)\xi(t-h_{k+1}) + 2\sum_{k=1}^{2}\xi^{\intercal}(t-h_{k})Z_{ik}\xi(t-h_{k+1}) \\
  & + 2\sum_{k=1}^{2}\xi^{\intercal}(t-\tau_{k}(t))\Big(T_{ik}-Z_{ik}^{\intercal}\Big)\xi(t-h_{k}) \triangleq \psi^{\intercal}(t)\widetilde{\Gamma}_{i}(t)\psi(t)\;,
\end{array}
\end{align*}
where $\psi(t)=\mbox{col}\big\{\xi(t),\xi(t-\tau_{1}(t)),\xi(t-\tau_{2}(t)),\xi(t-h_{1}),\xi(t-h_{2}),\xi(t-h_{3})\big\}$.
Note that the time derivative of $V_{i}(t)$ is bounded by a quadratic function in $\psi(t)$, \emph{i.e.},
\begin{align*}
\dot{V}_{i}(t)\leq \psi^{\intercal}(t)\widetilde{\Gamma}_{i}(t)\psi(t)\;,
\end{align*}
with
\begin{align*}
\widetilde{\Gamma}_{i}(t)=\lambda_{1}(t)\widetilde{\Gamma}_{i1} + \lambda_{2}(t)\widetilde{\Gamma}_{i2}\
\end{align*}
for all $i\in\{1,2\}$. Then, for two different modes, we form the following two matrices:
\begin{equation}
\renewcommand{\arraystretch}{1.15}
\setlength{\arraycolsep}{1.25pt}
\widetilde{\Gamma}_{1j}=
\left[\begin{array}{c:cc:ccc}
\Phi_{1} & P_{1}A_{\alpha_{1j}} & 0 & S_{11} & S_{12} & 0 \\ \hdashline
\star & -\Upsilon_{11}^{\mathcal{S}}  & 0 & \Upsilon_{11} ^{\intercal} & \Upsilon_{11} & 0 \\
\star & \star & -\Upsilon_{12}^{\mathcal{S}} & 0 & \Upsilon_{12}^{\intercal} & \Upsilon_{12} \\ \hdashline
\star & \star & \star & -\Xi_{11} & Z_{11} & 0 \\ 
\star & \star & \star & \star & -\Xi_{12} & Z_{12} \\
\star & \star & \star & \star & \star & -\Xi_{13}
\end{array}\right] + \phi_{1}^{\intercal}\sum_{k=1}^{2}\Big(h_{k}^{2}S_{1k}+\delta_{k}^{2}T_{1k}\Big)\phi_{1} \;,
\end{equation}
where $\Upsilon_{1j}\triangleq T_{1j}-Z_{1j}$ and $\phi_{1}=\big[~A_{\alpha}~A_{\alpha_{1j}}~0_{n\times 4n}~\big]$ for all $j\in\{1,2\}$, and
\begin{equation}
\renewcommand{\arraystretch}{1.15}
\setlength{\arraycolsep}{1.25pt}
\widetilde{\Gamma}_{2j}=
\left[\begin{array}{c:cc:ccc}
\Phi_{2} & 0 & P_{2}A_{\alpha_{2j}} & S_{21} & S_{22} & 0 \\ \hdashline
\star & -\Upsilon_{21}^{\mathcal{S}} & 0 & \Upsilon_{21}^{\intercal} & \Upsilon_{21} & 0 \\
\star & \star & -\Upsilon_{22}^{\mathcal{S}} & 0 & \Upsilon_{22}^{\intercal} & \Upsilon_{22} \\ \hdashline
\star & \star & \star & -\Xi_{21} & Z_{21} & 0 \\
\star & \star & \star & \star & -\Xi_{22} & Z_{22} \\
\star & \star & \star & \star & \star & -\Xi_{23}
\end{array}\right] +\phi_{2}^{\intercal}\sum_{k=1}^{2}\Big(h_{k}^{2}S_{2k}+\delta_{k}^{2}T_{2k}\Big)\phi_{2} \;,
\end{equation}
where $\Upsilon_{2j}\triangleq T_{2j}-Z_{2j}$ and $\phi_{2}=\big[~A_{\alpha}~0_{n}~A_{\alpha_{2j}}~0_{n\times 3n}~\big]$ for all $j\in\{1,2\}$. 

By applying the Schur complement twice to $\widetilde{\Gamma}_{ij}$ to form $\Gamma_{ij}$, we arrive at the equivalent condition:
\begin{equation}
\Gamma_{i}(t)=\lambda_{1}(t)\Gamma_{i1} + \lambda_{2}(t)\Gamma_{i2}< 0, \quad \forall i\in\{1,2\}\,.
\end{equation}
As argued above, the condition is satisfied for all $\lambda_{i}(t)$ if $\Gamma_{1i}$ and $\Gamma_{i2}$ are both negative definite.
By guaranteeing that $\Gamma_{i}(t)<0$, we ensure that the dynamics in each fixed mode is exponentially stable with decay rate $\alpha$. However, to guarantee stability for the switched system under the average dwell-time assumption, we also need to guarantee that
\begin{equation}
V_{i}(t)\leq \mu V_{j}(t)\,, \quad \forall i,j\in\{1,2\} \label{eq:swicthed_cond}
\end{equation}
for some $\mu>1$. Noting that $V_{i}(t)$ is linear in $P_{i}$, $Q_{ik}$, $R_{ik}$, $S_{ik}$ and $T_{ik}$,~\eqref{eq:swicthed_cond} is implied by the following conditions:
\begin{equation*}
P_{i}\leq \mu P_{j}\,,~Q_{ik}\leq \mu Q_{jk}\,,~R_{ik}\leq \mu R_{jk}\,,~S_{ik}\leq \mu S_{jk}\,,~T_{ik}\leq \mu T_{jk}
\end{equation*}
for all $i,j,k\in\{1,2\}$. This concludes the proof.\hfill$\square$

\begin{remark}
The analysis procedure extends immediately to the system with $M$ modes. However, the LMIs grow in both size and number. In contrast to the two-mode case, we need to check $2M$ LMIs (extensions of~\eqref{eq:thm1a},~\eqref{eq:thm1b}) whose dimensions are $2(2M+1)n\times 2(2M+1)n$, $M^{2}$ supplementary LMIs (\emph{e.g.},~\eqref{eq:thm1b}), and $M(4M^{2}-3M-1)$ additional LMIs (\emph{e.g.}, $P_{i}\leq\mu P_{j}$). The LMIs use $M(5M+1)$ matrix variables, each with $n(n+1)/2$ decision variables.
\end{remark}

\begin{prop}
A lower bound on the average dwell-time ensuring the global stability of switched delay system~\eqref{eq:SwitchedSys} is given by $\tau_a^{\circ}=\ln\mu/\alpha_{\circ}$ where $\alpha_{\circ}$ is the optimal value of the convex optimization problem
\begin{align}
\begin{cases}
\begin{array}{cl}
\underset{\substack{P_{i}>0,Q_{ik}>0,\\ R_{ik}>0,S_{ik}>0,T_{ik}>0}}{\mbox{maximize}} & \alpha \\ 
\mbox{subject~to} & 
\begin{array}{l}
\mathit{LMIs}~\eqref{eq:thm1a},~\eqref{eq:thm1b},~\mathit{and}~\eqref{eq:thm1c}, \\  P_{i}\leq\mu P_{j}, Q_{ik}\leq\mu Q_{jk}, R_{ik}\leq\mu R_{jk}, S_{ik}\leq\mu S_{jk}, T_{ik}\leq\mu T_{jk} \;\forall\; i,j,k\in \{1,2\} \;.
\end{array} 
\end{array}
\end{cases} \label{eq:optprob}
\end{align} 
\end{prop} 

Due to the presence of multiple product terms $\alpha P_{i}$ and $e^{\alpha h_{i+j-1}}A_{i}P_{i}$ in~\eqref{eq:thm1a} and \eqref{eq:thm1b}, the problem cannot be solved directly using semidefinite programming. However, the problem is easily seen to be quasi-convex. Hence, we can solve it by bisection in $\alpha$. Since the decay rate $\alpha$ is inversely proportional to $\tau_{a}$, this solution procedure gives us a lower bound on the allowable average dwell-time $\tau_{a}$. If we can guarantee that the average dwell-time between mode changes in the communication network is larger than this bound, then global stability of the closed-loop is guaranteed. 

\subsection{State-Feedback Controller Design}\label{sec:SwitchedSyssSynthesis}
In this section, we will extend our analysis conditions to mode-dependent state feedback synthesis for the supervisory control structure introduced in Section~\ref{sec:sysmodel}. More precisely, we consider a linear time-invariant plant
\begin{align*}
\dot{x}(t) &= Ax(t)+Bu(t)
\end{align*}
where the control input is a mode-dependent linear feedback of the delayed state vector, \emph{i.e},
\begin{align}
u(t) &= K_i x(t-\tau_i(t)) \label{eq:SwitchedSF}
\end{align}
when $\sigma(t)=i$ (and hence, $\tau_i(t)\in [h_i, h_i+1)$),\, $i\in {\mathcal M}$. The design problem is to find feedback gain matrices $K_i$ that ensure closed-loop stability for all switching signals in ${\mathcal S}[\tau_a]$. Clearly, this problem is closely related to the stability analysis problem considered in Section 3, since the supervisory control structure induces a switched linear system on the form~\eqref{eq:SwitchedSys} with $A_{\sigma(t)}=BK_{\sigma(t)}$. We have the following result:

\begin{thm}\label{thm:Switched_System_Stabilization}
For a given decay rate $\alpha>0$, there exists a state-feedback control of the form \eqref{eq:SwitchedSF} which exponentially stablizes system~\eqref{eq:SwitchedSys} over $\mathcal{S}[\tau_{a}]$ for time-varying delays $\tau_{i}(t)\in[h_{i},h_{i+1}),~\forall i\in\{1,2\}$ if there exist real constant matrices $\tilde{P}_{i}, \tilde{Q}_{ik}, \tilde{R}_{ik}, \tilde{S}_{ik}, \tilde{T}_{ik}\in\mathbb{S}^{n}_{++}~\forall i,k\in\{1,2\}$ and $\tilde{X}_{i}, \tilde{Z}_{ik}\in\mathbb{R}^{n\times n}~\forall i,k\in\{1,2\}$, and a constant scalar $\mu>1$ such that the LMIs
\begin{align}
\renewcommand{\arraystretch}{1.20}
\setlength{\arraycolsep}{1.6pt}
\left[\begin{array}{c:c:cc:ccc:c:cccc}
-\tilde{X}_{1}^{\mathcal{S}} & A_{\alpha}\tilde{X}_{1}+\tilde{P}_{1} & \varrho_{1j}B\tilde{Y}_{1} & 0 & 0 & 0 & 0 & \tilde{X}_{1} & h_{1}\tilde{S}_{11} & \delta_{1}\tilde{T}_{11} & h_{2}\tilde{S}_{12} & \delta_{2}\tilde{T}_{12} \\ \hdashline
\star & \sum_{k=1}^{2}\big(\tilde{Q}_{1k}+\tilde{R}_{1k}-\tilde{S}_{1k}\big)-\tilde{P}_{1} & 0 & 0 & \tilde{S}_{11} & \tilde{S}_{12} & 0 & 0 & 0 & 0 & 0 & 0 \\ \hdashline
\star & \star & -\tilde{\Upsilon}_{11}^{\mathcal{S}} & 0 & \tilde{\Upsilon}_{11}^{\intercal} & \tilde{\Upsilon}_{11} & 0 & 0 & 0 & 0 & 0 & 0 \\
\star & \star & \star & -\tilde{\Upsilon}_{12}^{\mathcal{S}} & 0 & \tilde{\Upsilon}_{12}^{\intercal} & \tilde{\Upsilon}_{12} & 0  & 0 & 0 & 0 & 0  \\ \hdashline
\star & \star & \star & \star & -\tilde{\Xi}_{11} & \tilde{Z}_{11} & 0 & 0 & 0 & 0 & 0 & 0 \\
\star & \star & \star & \star & \star & -\tilde{\Xi}_{12} & \tilde{Z}_{12} & 0 & 0 & 0 & 0 & 0 \\
\star & \star & \star & \star & \star & \star & -\tilde{\Xi}_{13} & 0 & 0 & 0 & 0 & 0  \\ \hdashline
\star & \star & \star & \star & \star &\star & \star & -\tilde{P}_{1} & -h_{1}\tilde{S}_{11} & -\delta_{1}\tilde{T}_{11} & -h_{2}\tilde{S}_{12} & -\delta_{2}\tilde{T}_{12} \\ \hdashline
\star & \star & \star & \star & \star &\star & \star & \star & -\tilde{S}_{11} & 0 & 0 & 0\\
\star & \star & \star & \star & \star &\star & \star & \star & \star & -\tilde{T}_{11} & 0 & 0\\
\star & \star & \star & \star & \star &\star & \star & \star & \star & \star & -\tilde{S}_{12} & 0 \\
\star & \star & \star & \star & \star &\star & \star & \star & \star & \star & \star & -\tilde{T}_{12} \\
\end{array}\right] < 0 \label{eq:thm2c} \\
\renewcommand{\arraystretch}{1.20}
\setlength{\arraycolsep}{1.6pt}
\left[\begin{array}{c:c:cc:ccc:c:cccc}
-\tilde{X}_{2}^{\mathcal{S}} & A_{\alpha}\tilde{X}_{2}+\tilde{P}_{2} & 0 & \varrho_{2j}B\tilde{Y}_{2} & 0 & 0 & 0 & \tilde{X}_{2} & h_{1}\tilde{S}_{21} & \delta_{1}\tilde{T}_{21} & h_{2}\tilde{S}_{22} & \delta_{2}\tilde{T}_{22} \\ \hdashline
\star & \sum_{k=1}^{2}\big(\tilde{Q}_{2k}+\tilde{R}_{2k}-\tilde{S}_{2k}\big)-\tilde{P}_{2} & 0 & 0 & \tilde{S}_{21} & \tilde{S}_{22} & 0 & 0 & 0 & 0 & 0 & 0 \\ \hdashline
\star & \star & -\tilde{\Upsilon}_{21}^{\mathcal{S}} & 0 & \tilde{\Upsilon}_{21}^{\intercal} & \tilde{\Upsilon}_{21} & 0 & 0 & 0 & 0 & 0 & 0 \\
\star & \star & \star & -\tilde{\Upsilon}_{22}^{\mathcal{S}} & 0 & \tilde{\Upsilon}_{22}^{\intercal} & \tilde{\Upsilon}_{22} & 0  & 0 & 0 & 0 & 0  \\ \hdashline
\star & \star & \star & \star & -\tilde{\Xi}_{21} & \tilde{Z}_{21} & 0 & 0 & 0 & 0 & 0 & 0 \\
\star & \star & \star & \star & \star & -\tilde{\Xi}_{22} & \tilde{Z}_{22} & 0 & 0 & 0 & 0 & 0 \\
\star & \star & \star & \star & \star & \star & -\tilde{\Xi}_{23} & 0 & 0 & 0 & 0 & 0  \\ \hdashline
\star & \star & \star & \star & \star &\star & \star & -\tilde{P}_{2} & -h_{1}\tilde{S}_{21} & -\delta_{1}\tilde{T}_{21} & -h_{2}\tilde{S}_{22} & -\delta_{2}\tilde{T}_{22} \\ \hdashline
\star & \star & \star & \star & \star &\star & \star & \star & -\tilde{S}_{21} & 0 & 0 & 0\\
\star & \star & \star & \star & \star &\star & \star & \star & \star & -\tilde{T}_{21} & 0 & 0\\
\star & \star & \star & \star & \star &\star & \star & \star & \star & \star & -\tilde{S}_{22} & 0 \\
\star & \star & \star & \star & \star &\star & \star & \star & \star & \star & \star & -\tilde{T}_{22} \\
\end{array}\right] < 0 \label{eq:thm2d} 
\end{align}
\begin{align}
\begin{bmatrix}
\tilde{T}_{ik} & \tilde{Z}_{ik} \\ \star & \tilde{T}_{ik}
\end{bmatrix} \geq 0 \label{eq:thm2e} 
\end{align}
where $\tilde{\Xi}_{i1}=\tilde{Q}_{i1}+\tilde{T}_{i1}+\tilde{S}_{i1}$,~$\tilde{\Xi}_{i2}=\tilde{Q}_{i2}+\tilde{S}_{i2}+\sum_{k=1}^{2}\tilde{T}_{ik}+\tilde{R}_{i1}$~$\tilde{\Xi}_{i3}=\tilde{R}_{i2}+\tilde{T}_{i2}$, $\tilde{\Upsilon}_{ik}=\tilde{T}_{ik}-\tilde{Z}_{ik}$, and $\tilde{P}_{i}\leq\mu\tilde{P}_{j}$, $\tilde{Q}_{ik}\leq\mu\tilde{Q}_{jk}$, $\tilde{R}_{ik}\leq\mu\tilde{R}_{jk}$, $\tilde{S}_{ik}\leq\mu\tilde{S}_{jk}$ and $\tilde{T}_{ik}\leq\mu\tilde{T}_{jk}~\forall i,j,k\in\{1,2\}$ are feasible. A stabilizing control law is given by~\eqref{eq:SwitchedSF} with gain $K_{i}=\tilde{Y}_{i}\tilde{X}_{i}^{-1}$ for all $i\in\{1,2\}$.
\end{thm}

\noindent\textbf{Proof:} The structure of~\eqref{eq:thm1a} and~\eqref{eq:thm1b} is not suitable for the synthesis of a state-feedback controller due to the presence of multiple product terms $A_{\alpha}S_{ik}$, $A_{\alpha}T_{ik}$, $A_{\alpha_{ij}}S_{ik}$ and $A_{\alpha_{ij}}T_{ik}$. These product terms prevent finding a linearizing change of variable even after congruence transformation. Instead, we will use the relaxation term introduced in Briat~\emph{et.~al}~\cite{BSL:10} to decouple the products at the expense of an increased conservatism. Denote~\eqref{eq:thm2c} and~\eqref{eq:thm2d} by $\Theta_{1j}$ and $\Theta_{2j}$, respectively. Then we prove that $\Theta_{ij}< 0~\forall i\in\{1,2\}$ implies the feasibility of~\eqref{eq:thm1a} and~\eqref{eq:thm1b}. Note that $\Theta_{ij}$ can be decomposed as
\begin{align*}
\Theta_{ij}=\Theta_{ij}\vert_{X_{i}=0}+U_{i}^{\intercal}X_{i}V_{i}+V_{i}^{\intercal}X_{i}^{\intercal}U_{i}< 0, \quad \forall i\in\{1,2\}
\end{align*}\label{eq:Project}
where $U_{1}=\big[-I_{n}~A_{\alpha}~A_{\alpha_{1j}}~0_{n\times 4n}~I_{n}~0_{n\times 4n}\big]$, $V_{1}=\big[I_{n}~0_{n\times 11n}\big]$, $U_{2}=\big[-I_{n}~A_{\alpha}~0_{n}~A_{\alpha_{2j}}~0_{n\times 3n}~I_{n}~0_{n\times 4n}\big]$ and $V_{2}=\big[I_{n}~0_{n\times 11n}\big]$. Then invoking the projection lemma~\cite{GaA:94}, the feasibility of $\Theta_{ij}< 0$ implies the feasibility of the LMIs
\begin{align}
\mathcal{N}_{U_{i}}^{T}\Theta_{ij}|_{X_{i}=0}\mathcal{N}_{U_{i}}< 0 \label{eq:adjLMIa} \\
\mathcal{N}_{V_{i}}^{T}\Theta_{ij}|_{X_{i}=0}\mathcal{N}_{V_{i}}< 0 \label{eq:adjLMIb}
\end{align}
where $\mathcal{N}_{U_{i}}$ and $\mathcal{N}_{V_{i}}$ are basis of the null space of $U_{i}$ and $V_{i}$, respectively. After some tedious calculations, we can show that LMIs~\eqref{eq:thm2a} and~\eqref{eq:thm2b} are equivalent to~\eqref{eq:thm1a} and~\eqref{eq:thm1b} showing that $\Theta_{ij}< 0~\forall i\in\{1,2\}$ implies the feasibility of~\eqref{eq:thm1a} and~\eqref{eq:thm1b}. Moreover, LMI~\eqref{eq:adjLMIb} characterizes the conservatism of the relaxation.

\begin{align}
\renewcommand{\arraystretch}{1.2}
\setlength{\arraycolsep}{1.6pt}
\left[\begin{array}{c:c:cc:ccc:c:cccc}
-X_{1}^{\mathcal{S}} & X_{1}^{\intercal}A_{\alpha}+P_{1} & X_{1}^{\intercal}A_{\alpha_{1j}} & 0 & 0 & 0 & 0 & X_{1}^{\intercal} & h_{1}S_{11} & \delta_{1}T_{11} & h_{2}S_{12} & \delta_{2}T_{12} \\ \hdashline
\star & \sum_{k=1}^{2}\big(Q_{1k}+R_{1k}-S_{1k}\big)-P_{1} & 0 & 0 & S_{11} & S_{12} & 0 & 0 & 0 & 0 & 0 & 0 \\ \hdashline
\star & \star & -\Upsilon_{11}^{\mathcal{S}} & 0 & \Upsilon_{11}^{\intercal} & \Upsilon_{11} & 0 & 0 & 0 & 0 & 0 & 0 \\ 
\star & \star & \star & -\Upsilon_{12}^{\mathcal{S}} & 0 & \Upsilon_{12}^{\intercal} & \Upsilon_{12} & 0  & 0 & 0 & 0 & 0  \\ \hdashline
\star & \star & \star & \star & -\Xi_{11} & Z_{11} & 0 & 0 & 0 & 0 & 0 & 0 \\
\star & \star & \star & \star & \star & -\Xi_{12} & Z_{12} & 0 & 0 & 0 & 0 & 0 \\
\star & \star & \star & \star & \star & \star & -\Xi_{13} & 0 & 0 & 0 & 0 & 0  \\ \hdashline
\star & \star & \star & \star & \star &\star & \star & -P_{1} & -h_{1}S_{11} & -\delta_{1}T_{11} & -h_{2}S_{12} & -\delta_{2}T_{12} \\ \hdashline
\star & \star & \star & \star & \star &\star & \star & \star & -S_{11}  & 0 & 0 & 0\\
\star & \star & \star & \star & \star &\star & \star & \star & \star & -T_{11} & 0 & 0\\
\star & \star & \star & \star & \star &\star & \star & \star & \star & \star & -S_{12} & 0 \\
\star & \star & \star & \star & \star &\star & \star & \star & \star & \star & \star & -T_{12} \\
\end{array}\right] < 0\label{eq:thm2a} \\
\renewcommand{\arraystretch}{1.2}
\setlength{\arraycolsep}{1.6pt}
\left[\begin{array}{c:c:cc:ccc:c:cccc}
-X_{2}^{\mathcal{S}} & X_{2}^{\intercal}A_{\alpha}+P_{2} & 0 & X_{2}^{\intercal}A_{\alpha_{2j}} & 0 & 0 & 0 & X_{2}^{\intercal} & h_{1}S_{21} & \delta_{1}T_{21} & h_{2}S_{22} & \delta_{2}T_{22} \\ \hdashline
\star & \sum_{k=1}^{2}\big(Q_{2k}+R_{2k}-S_{2k}\big)-P_{2} & 0 & 0 & S_{21} & S_{22} & 0 & 0 & 0 & 0 & 0 & 0 \\ \hdashline
\star & \star & -\Upsilon_{21}^{\mathcal{S}} & 0 & \Upsilon_{21}^{\intercal} & \Upsilon_{21} & 0 & 0 & 0 & 0 & 0 & 0 \\
\star & \star & \star & -\Upsilon_{22}^{\mathcal{S}} & 0 & \Upsilon_{22}^{\intercal} & \Upsilon_{22} & 0  & 0 & 0 & 0 & 0  \\ \hdashline
\star & \star & \star & \star & -\Xi_{21} & Z_{21} & 0 & 0 & 0 & 0 & 0 & n   0 \\
\star & \star & \star & \star & \star & -\Xi_{22} & Z_{22} & 0 & 0 & 0 & 0 & 0 \\
\star & \star & \star & \star & \star & \star & -\Xi_{23} & 0 & 0 & 0 & 0 & 0  \\ \hdashline
\star & \star & \star & \star & \star &\star & \star & -P_{2} & -h_{1}S_{21} & -\delta_{1}T_{21} & -h_{2}S_{22} & -\delta_{2}T_{22} \\ \hdashline
\star & \star & \star & \star & \star &\star & \star & \star & -S_{21}  & 0 & 0 & 0\\
\star & \star & \star & \star & \star &\star & \star & \star & \star & -T_{21} & 0 & 0\\
\star & \star & \star & \star & \star &\star & \star & \star & \star & \star & -S_{22} & 0 \\
\star & \star & \star & \star & \star &\star & \star & \star & \star & \star & \star & -T_{22} \\
\end{array}\right] < 0\label{eq:thm2b} 
\end{align}

Since LMIs~\eqref{eq:thm2a} and~\eqref{eq:thm2b} do not include any multiple product, it can easily be used for controller design. Hence, it is possible to use congruence transformations and change of variables so as to design the state-feedback controller. Performing a congruence transformation with respect to matrix $I_{12n}\otimes X^{-1}$ and applying the following linearizing change of variables $\tilde{X}_{i}\triangleq X_{i}^{-1},~\tilde{P}_{i}\triangleq \tilde{X}_{i}^{\intercal}P_{i}\tilde{X}_{i},~\tilde{Q}_{ik}\triangleq \tilde{X}_{i}^{\intercal}Q_{ik}\tilde{X}_{i},~\tilde{R}_{ik}\triangleq \tilde{X}_{i}^{\intercal}R_{ik}\tilde{X}_{i},~\tilde{S}_{ik}\triangleq \tilde{X}_{i}^{\intercal}S_{ik}\tilde{X}_{i},~\tilde{T}_{ik}\triangleq \tilde{X}_{i}^{\intercal}T_{ik}\tilde{X}_{i},~\tilde{\Xi}_{i1}\triangleq\tilde{X}_{i}^{\intercal}\Xi_{i1}\tilde{X}_{i},~\tilde{\Xi}_{i2}\triangleq\tilde{X}_{i}^{\intercal} \Xi_{i2}\tilde{X}_{i},~\tilde{\Xi}_{i3}\triangleq \tilde{X}_{i}^{\intercal}\Xi_{i3}\tilde{X}_{i},~\tilde{Z}_{ik}\triangleq \tilde{X}_{i}^{\intercal}Z_{ik}\tilde{X}_{i}$ and $\tilde{Y}_{i}=K_{i}\tilde{X}_{i},~\forall i,k\in\{1,2\}$ yields LMI~\eqref{eq:thm2c} and~\eqref{eq:thm2d}. \hfill $\square$

\begin{remark}
The synthesis procedure readily extends to systems with $M$ modes, yet both the size and the number of LMIs increase. Specifically, we need to check $2M$ LMIs (extensions of~\eqref{eq:thm2c},~\eqref{eq:thm2d}) whose dimensions are $2(2M+2)n\times 2(2M+2)n$ and $M(5M^{2}-3M-1)$ additional LMIs (\emph{e.g.},~\eqref{eq:thm2e}, $\tilde{P}_{i}\leq\mu \tilde{P}_{j}$). In total, these LMIs comprise $M(5M+3)$ matrix variables, each of which has $n(n+1)/2$ decision variables.
\end{remark}

\section{Stochastic Switched Systems}\label{sec:StocSwitchedSys} 

Our deterministic modeling framework has several advantages: it allows to model long time-delays, is able to account for mode-dependent delay bounds and admits a convex formulation of the (mode-dependent) state feedback synthesis problem. However, it also has a disadvantage in that it does not allow to account for more detailed knowledge about the evolution of the delay mode beyond the average dwell-time.  It is therefore interesting to derive similar results when the delay mode varies according to a Markov chain, cf. ~\cite{BLL:10,WCW:06,BeB:98,BoL:02}. Such results will be developed next.

\subsection{System Model}\label{subsec:StochSystemModel}

Let us consider a dynamical system in a probability space $(\Omega,\mathcal{F},\mathbf{P})$, where $\Omega$ is the sample space, $\mathcal{F}$ is the $\sigma$-algebra of subsets of the sample space and $\mathbf{P}$ is the probability measure on $\mathcal{F}$. Over this probability space, we consider the following class of linear stochastic systems with Markovian jump parameters and mode-dependent time delays
\begin{align}
\begin{array}{rll}
\Sigma_{2}: & \dot{x}(t) = Ax(t) + A_{r(t)}x\big(t-\tau_{r(t)}(t)\big)\;, & \forall t\in \mathbb{R}_{\geq 0}\;, \\
 & x(t) = \varphi(t)\;, & \forall t\in [-h_{M+1},0]\;, \label{eq:MJLS}
\end{array}
\end{align} 
Here, $x(t)\in\mathbb{R}^{n}$ is the state, $A\in\mathbb{R}^{n\times n}$ and $A_{r(t)}\in\mathbb{R}^{n\times n}$ are the known system matrices while
$\big\{ r_{t}, t\in\mathbb{R}_{\geq 0} \big\}$ is a homogeneous, finite-state Markovian process with right continuous trajectories and taking values in the finite set $\mathcal{M}=\{1,\cdots,M\}$. The Markov process describes the switching between the different modes and its evolution is governed by the following transition probabilities 
\begin{equation*}
\mathbf{P}\big[ r_{t+\Delta}=j\;\vert \; r_{t}=i \big] = 
\begin{cases} 
\pi_{ij}\Delta + \mathit{o}(\Delta) & \text{if}~i\neq j\;, \\ 
1 + \pi_{ii}\Delta + \mathit{o}(\Delta), & \text{if}~i=j\;,
\end{cases} 
\end{equation*} 
where $\pi_{ij}$ is the transition rate from mode $i$ to $j$ with $\pi_{ij}\geq 0$ when $i\neq j$ and $\pi_{ii}=-\sum_{j=1,j\neq i}^{N}\pi_{ij}$ and $\mathit{o}(\Delta)$ is such that $\lim_{\Delta\rightarrow 0}\frac{\mathit{o}(\Delta)}{\Delta}=0$. Furthermore, $\tau_{r(t)}(t)$ is the time-varying stochastic delay function satisfying 
\begin{equation*}
h_{1}\leq \underline{h}_{r(t)} \leq \tau_{r(t)}(t) \leq \overline{h}_{r(t)} \leq h_{M+1}\;.
\end{equation*}
Finally, $\varphi(t)$ is a vector-valued initial continuous function defined on the interval $[-h_{M+1} 0]$, and $r_{0}\in\mathcal{M}$ are the initial conditions of the continuous state and the mode. 

\begin{defi}
The Markovian jump system~\eqref{eq:MJLS} is exponentially mean-square stable if there exist positive constants $\alpha$ and $\gamma$ such that 
\begin{equation*}
\mathbb{E}\Big[ \parallel x(t) \parallel^{2} \big\vert\; \varphi(t_{0}), r_{t_{0}} \Big] \leq\; \gamma\parallel x(t_{0},r_{t_{0}}) \parallel^{2}e^{-\alpha (t-t_{0})}
\end{equation*}
holds for any finite $\varphi(t_{0})\in\mathbb{R}^{n}$ defined on $[-h_{M+1},0]$ and any initial mode $r_{t_{0}}\in\mathcal{M}$. 
\end{defi}

\subsection{Exponential Stability Analysis Using Stochastic Lyapunov-Krasovskii Functionals}\label{subsec:StocStability}

In this subsection, we analyze the exponential stability of the Markovian jump linear system~\eqref{eq:MJLS} using a similar approach to what we developed for the switched delay case.
To portray the convergence rate of the system~\eqref{eq:MJLS}, we thus consider the change of variables $\xi(t)\triangleq e^{\alpha t}x(t)$ and find
\begin{align}
\dot{\xi}(t) = &\; (\alpha I_{n}+A)\xi(t) +e^{\alpha\tau_{r(t)}(t)}A_{r(t)}\xi(t-\tau_{r(t)}(t)) \;, 
\label{eq:MJLS_exp}
\end{align}
where $\tau_{r(t)}\in\big[\underline{h}_{r(t)}, \overline{h}_{r(t)}\big)$. For each $r(t)=i, \forall i\in\mathcal{M}$, we rewrite~\eqref{eq:MJLS_exp} as
\begin{equation}
\dot{\xi}(t) = (\alpha I_{n}+A)\xi(t) +e^{\alpha\tau_{i}(t)}A_{i}\xi(t-\tau_{i}(t)) \;. 
\label{eq:MJLS_exp2}
\end{equation}
Using the same polytopic approach as in Section~2, we express $e^{\alpha\tau_{i}(t)}$ as a convex combination of its mode-dependent bounds:
\begin{equation*}
e^{\alpha\tau_{i}(t)}=\lambda_{1}(t)e^{\alpha h_{i}}+\lambda_{2}(t)e^{\alpha h_{i+1}}\,,\quad \forall i\in\mathcal{M}
\end{equation*}
where $\lambda_{1}(t),\lambda_{2}(t)\in\mathbb{R}_{\geq 0}$ and $\lambda_{1}(t)+\lambda_{2}(t)=1,~\forall t\in\mathbb{R}_{\geq 0}$. Thus, the stochastic switched system~\eqref{eq:MJLS} can be defined, for each $r(t)=i, \forall i\in\mathcal{M}$, as
\begin{align}
\dot{\xi}(t)=A_{\alpha}\xi(t)+\sum_{j=1}^{2}\lambda_{j}(t)A_{\alpha_{ij}}\xi\big(t-\tau_{i}(t)\big)\label{SwitchedSysTV2}
\end{align}
where $A_{\alpha}\triangleq (\alpha I_{n}+A)$ and $A_{\alpha_{ij}}\triangleq\varrho_{ij}A_{i}$~with~$\varrho_{ij}\triangleq e^{\alpha h_{i+j-1}}$ when $\tau_{i}(t)\in [h_{i},h_{i+1})\;,~\forall i\in\mathcal{M}$.

\begin{thm}
The Markovian jump linear system~\eqref{eq:MJLS} is exponentially mean-square stable with a given decay rate $\alpha >0$ for randomly varying delays $\tau_{i}\in [h_{i},h_{i+1}),~\forall i\in\{1,2\}$ if there exist matrices $P_{i}, Q_{i}, R_{i}\in\mathbb{S}_{++}^{n},~\forall i\in\{1,2\}$, $S, T, \mathcal{Q}, \mathcal{R}\in\mathbb{S}_{++}^{n}$ and $Z\in\mathbb{R}^{n\times n}$ such that the following LMIs hold for $i, j \in \{1,2\}$
\begin{align}
\begin{bmatrix} T & Z \\ Z^{\intercal} & T\end{bmatrix} \geq 0\;, \qquad \sum_{j=1}^{2}\pi_{ij}Q_{j}\leq\mathcal{Q}\;, \qquad \sum_{j=1}^{2}\pi_{ij}R_{j}\leq\mathcal{R}\;, \label{eq:MJLS_Stability_add_LMI} \\
\renewcommand{\arraystretch}{1.4}
\setlength{\arraycolsep}{1.6pt}
\left[\begin{array}{c:c:cc:cc}
\Phi_{i} & P_{i}A_{\alpha_{ij}} & S & 0 & \sqrt{\epsilon_{1,i}}A_{\alpha}^{\intercal}S & \sqrt{\epsilon_{2,i}}A_{\alpha}^{\intercal}T  \\ \hdashline
\star & -\Upsilon^{\mathcal{S}} & \Upsilon^{\intercal} & \Upsilon & \sqrt{\epsilon_{1,i}}A_{\alpha_{ij}}^{\intercal}S & \sqrt{\epsilon_{2,i}}A_{\alpha_{ij}}^{\intercal}T \\ \hdashline
\star & \star & -(Q_{i}+S+T) & Z & 0 & 0 \\ 
\star & \star & \star & -(R_{i}+T) & 0 & 0 \\ \hdashline
\star & \star & \star & \star & -S & 0 \\
\star & \star & \star & \star & \star & -T
\end{array}\right] < 0\;, \label{eq:MJLS_Stability_LMI}
\end{align}
where $\Phi_{i}\triangleq A_{\alpha}^{\intercal}P_{i}+P_{i}A_{\alpha}+\sum_{j=1}^{2}\pi_{ij}P_{j} + \big(Q_{i} + h_{2}\mathcal{Q} + \delta_{1}Q_{\kappa}\big) + \big(R_{i} + h_{3}\mathcal{R} + \delta_{2}R_{\kappa}\big) - S$, $\epsilon_{1,i}\triangleq h_{i}^{2} + \eta\frac{h_{2}^{3}-h_{1}^{3}}{2}$ and $\epsilon_{2,i}\triangleq \delta_{i}^{2} + \eta\delta_{\max}\frac{h_{3}^{2}-h_{1}^{2}}{2}$ with $\eta\triangleq \max\vert\pi_{ii}\vert$, $\kappa\triangleq\mathrm{argmax}\vert\pi_{ii}\vert$ and $\delta_{\max}=\max\vert h_{i+1}-h_{i} \vert,~\forall i\in\{1,2\}$, and $\Upsilon = T - Z$.
\end{thm}

\noindent\textbf{Proof.} We define a stochastic Lyapunov-Krasovskii functional $V:\mathbb{R}^{n}\times\mathcal{M}\rightarrow\mathbb{R}_{\geq 0}$ candidate for the system~\eqref{eq:MJLS} as 
\begin{equation}
V(\xi_{t},r_{t}) = \underbrace{\xi^{\intercal}(t)P_{r(t)}\xi(t)}_{V_{1}(\xi_{t},r_{t})} + \sum_{k=2}^{5}V_{k}(\xi_{t},r_{t}) \;,
\end{equation}
where
\begin{align*} 
V_{2}(\xi_{t},r_{t}) = & \int_{t-\underline{h}_{r(t)}}^{t} \xi^{\intercal}(s)Q_{r(t)}\xi(s)ds + \int_{-h_{2}}^{0} \int_{t+s}^{t} \xi^{\intercal}(\theta)\mathcal{Q}\xi(\theta)d\theta ds + \eta\int_{-h_{2}}^{-h_{1}} \int_{t+s}^{t} \xi^{\intercal}(\theta)Q_{\kappa}\xi(\theta)d\theta ds \\
V_{3}(\xi_{t},r_{t}) = & \int_{t-\overline{h}_{r(t)}}^{t}\xi^{\intercal}(s)R_{r(t)}\xi(s)ds + \int_{-h_{3}}^{0} \int_{t+s}^{t} \xi^{\intercal}(\theta)\mathcal{R}\xi(\theta)d\theta ds + \eta\int_{-h_{3}}^{-h_{2}} \int_{t+s}^{t} \xi^{\intercal}(\theta)R_{\kappa}\xi(\theta)d\theta ds \\ 
V_{4}(\xi_{t},r_{t}) = &\; \underline{h}_{r(t)} \int_{-\underline{h}_{r(t)}}^{0} \int_{t+s}^{t} \dot{\xi}^{\intercal}(\theta) S \dot{\xi}(\theta)d\theta ds + \eta h_{2}\int_{-h_{2}}^{-h_{1}} \int_{s}^{0} \int_{t+\theta}^{t} \dot{\xi}^{\intercal}(\upsilon) S\dot{\xi}(\upsilon) d\upsilon d\theta ds \\
& \; + \eta\delta_{1}\int_{-h_{1}}^{0} \int_{s}^{0} \int_{t+\theta}^{t} \dot{\xi}^{\intercal}(\upsilon) S\dot{\xi}(\upsilon) d\upsilon d\theta ds\\
V_{5}(\xi_{t},r_{t}) = &\; \underbrace{\big(\overline{h}_{r(t)}-\underline{h}_{r(t)}\big)}_{\delta_{r(t)}}\int_{-\overline{h}_{r(t)}}^{-\underline{h}_{r(t)}} \int_{t+s}^{t} \dot{\xi}^{\intercal}(\theta) T\dot{\xi}(\theta)d\theta ds + \eta\delta_{\max}\int_{-h_{3}}^{-h_{1}}\int_{s}^{0}\int_{t+\theta}^{t}\dot{\xi}^{\intercal}(\upsilon)T\dot{\xi}(\upsilon)d\upsilon d\theta ds
\end{align*}
with $\eta\triangleq\max{\vert\pi_{ii}\vert}$, $\kappa\triangleq\mathrm{argmax}\vert\pi_{ii}\vert$, and $\delta_{\max}\triangleq\max\vert h_{i+1}-h_{i}\vert,~\forall i\in\mathcal{M}$.

\begin{defi}[\O ksendal~\cite{Oks:02}]
The weak infinitesimal operator $\mathcal{A}$ of the Markovian process $\big\{ \big(x(t),r_{t}\big),t\geq 0 \big\}$ is defined by
\begin{align*} 
\mathcal{A}V\big(x(t),r_{t}\big) = \lim_{\Delta\rightarrow 0} \frac{\mathbb{E}\Big[V\big(x(t+\Delta),r_{t+\Delta}\big)\big\vert \mathcal{F}_{t}\Big] - V\big(x(t),r_{t}\big)}{\Delta} \;,
\end{align*} 
where $\mathcal{F}_{t}=\sigma\big((x(t),r_{t}), t\geq 0\big)$.
\end{defi}

Straightforward but tedious calculations yield that, for each $r_{t}=i$, $i\in\mathcal{M}$, along solutions of~\eqref{eq:MJLS}, we have
\begin{align}
\mathcal{A}V_{1} =&\; \xi^{\intercal}(t)\Bigg[A_{\alpha}^{\intercal}P_{i}+P_{i}A_{\alpha}+\sum_{j=1}^{2}\pi_{ij}P_{j}\Bigg]\xi(t) + 2\xi^{\intercal}(t)P_{i}\sum_{j=1}^{2}(\lambda_{j}(t)A_{\alpha_{ij}})\xi\big(t-\tau_{i}(t)\big) \label{eq:Vdot1} \\
\mathcal{A}V_{2} =&\; \xi^{\intercal}(t)Q_{i}\xi(t) - \xi^{\intercal}(t-h_{i})Q_{i}\xi(t-h_{i}) + \sum_{j=1}^{2}\pi_{ij}\int_{t-h_{j}}^{t} \xi^{\intercal}(s)Q_{j}\xi(s)ds + h_{2} \xi^{\intercal}(t)\mathcal{Q}\xi(t) + \delta_{1}\eta\xi^{\intercal}(t)Q_{\kappa}\xi(t) \nonumber\\ 
&\; - \Bigg[ \int_{t-h_{2}}^{t} \xi^{\intercal}(s)\mathcal{Q}\xi(s)ds + \eta\int_{t-h_{2}}^{t-h_{1}} \xi^{\intercal}(s)Q_{\kappa}\xi(s)ds \Bigg] \label{eq:Vdot2} \\
\mathcal{A}V_{3} =&\; \xi^{\intercal}(t)R_{i}\xi(t) - \xi^{\intercal}(t-h_{i+1})R_{i}\xi(t-h_{i+1}) + \sum_{j=1}^{2}\pi_{ij}\int_{t-h_{j+1}}^{t} \xi^{\intercal}(s)R_{j}\xi(s)ds + h_{3}\xi^{\intercal}(t)\mathcal{R}\xi(t) + \delta_{2}\eta\xi^{\intercal}(t)R_{\kappa}\xi(t) \nonumber\\
&\;  - \Bigg[ \int_{t-h_{3}}^{t} \xi^{\intercal}(s)\mathcal{R}\xi(s)ds + \eta\int_{t-h_{3}}^{t-h_{2}} \xi^{\intercal}(s)R_{\kappa}\xi(s)ds \Bigg] \label{eq:Vdot3} \\
\mathcal{A}V_{4} =&\; h_{i}^{2}~\dot{\xi}^{\intercal}(t)S\dot{\xi}(t) - h_{i}\int_{t-h_{i}}^{t}\dot{\xi}^{\intercal}(s)S\dot{\xi}(s)ds + \sum_{j=1}^{2}\pi_{ij}~h_{j} \int_{-h_{j}}^{0} \int_{t+s}^{t} \dot{\xi}^{\intercal}(\theta) S \dot{\xi}(\theta)d\theta ds \nonumber\\ 
&\; + \eta\frac{h_{2}^{3}-h_{1}^{3}}{2}\dot{\xi}^{\intercal}(t)S\dot{\xi}(t) - \eta\Bigg[ h_{2}\int_{-h_{2}}^{-h_{1}}\int_{t+s}^{t}\dot{\xi}^{\intercal}(\theta)S\dot{\xi}(\theta)d\theta ds + (h_{2}-h_{1})\int_{-h_{1}}^{0}\int_{t+s}^{t}\dot{\xi}^{\intercal}(\theta)S\dot{\xi}(\theta)d\theta ds \Bigg] \label{eq:Vdot4} \\ 
\mathcal{A}V_{5} = &\; \delta_{i}^{2}\dot{\xi}^{\intercal}(t)T\dot{\xi}(t) - \delta_{i}\int_{t-h_{i+1}}^{t-h_{i}}\dot{\xi}^{\intercal}(s)T\dot{\xi}(s)ds + \sum_{j=1}^{2}\pi_{ij}\;\delta_{j}\int_{-h_{j+1}}^{-h_{j}}\int_{t+s}^{t}\dot{\xi}^{\intercal}(\theta)T\dot{\xi}(\theta)d\theta ds \nonumber\\
&\; + \eta\;\delta_{\max}\frac{h_{3}^{2}-h_{1}^{2}}{2}\dot{\xi}^{\intercal}(t)T\dot{\xi}(t) - \eta\;\delta_{\max}\int_{-h_{3}}^{-h_{1}}\int_{t+s}^{t}\dot{\xi}^{\intercal}(\theta)T\dot{\xi}(\theta)d\theta ds \label{eq:Vdot5} \;.
\end{align}

Similar to~Section~\ref{subsec:DeterStability}, we bound the integral terms $\int_{t-h_{i}}^{t}h_{i}\dot{\xi}^{\intercal}(s)S\dot{\xi}(s)ds$ and $\int_{t-h_{i+1}}^{t-h_{i}}\delta_{i}\dot{\xi}^{\intercal}(s)T\dot{\xi}(s)ds$ that appear in the preceding equalities as follows:
\begin{align}
-\int_{t-h_{i}}^{t}h_{i}\dot{\xi}^{\intercal}(s)S\dot{\xi}(s)ds &\leq \; -\big[ \xi(t) - \xi(t-h_{i}) \big]^{\intercal} S \big[ \xi(t) - \xi(t-h_{i}) \big]\;,  \label{eq:Jensen2}\\
-\int_{t-h_{i+1}}^{t-h_{i}}\delta_{i}\dot{\xi}^{\intercal}(s) T\dot{\xi}(s)ds &\leq \; -
\begin{bmatrix} \xi(t-h_{i})-\xi(t-\tau_{i}(t)) \\ \xi(t-\tau_{i}(t))-\xi(t-h_{i+1}) \end{bmatrix}^{\intercal} 
\begin{bmatrix} T & Z \\ \star & T \end{bmatrix}
\begin{bmatrix} \xi(t-h_{i})-\xi(t-\tau_{i}(t)) \\ \xi(t-\tau_{i}(t))-\xi(t-h_{i+1}) \end{bmatrix} \;, \label{eq:Reciprocal2}
\end{align}
where $\Bigl[\begin{smallmatrix} T & Z \\ \star & T \end{smallmatrix} \Bigr]\geq 0$ holds. 

In addition, for the stochastic formulation, we need to upper bound a number of additional integrals. We do so by noting $\pi_{ij}\geq 0$ for $j\neq i$ and $\pi_{ii}\leq 0$, and that
\begin{align}
\renewcommand{\arraystretch}{2}
\setlength{\arraycolsep}{1.5pt}
\begin{array}{rll}
\displaystyle{\sum_{j=1}^{M}\pi_{ij}\int_{t-h_{j}}^{t}\xi^{\intercal}(s)Q_{j}\xi(s)ds} & = \displaystyle{\sum_{j\neq i}^{M}\pi_{ij}\int_{t-h_{j}}^{t}\xi^{\intercal}(s)Q_{j}\xi(s)ds} & +\; \displaystyle{\pi_{ii}\int_{t-h_{i}}^{t}\xi^{\intercal}(s)Q_{i}\xi(s)ds} \\
																										  & \leq \displaystyle{\int_{t-h_{M}}^{t}\xi^{\intercal}(s)\Bigg(\sum_{j\neq i}^{M}\pi_{ij}Q_{j}\Bigg)\xi(s)ds} & +\; \displaystyle{\pi_{ii}\int_{t-h_{1}}^{t}\xi^{\intercal}(s)Q_{i}\xi(s)ds} \\
																										  & = \displaystyle{\int_{t-h_{M}}^{t}\xi^{\intercal}(s)\Big[\mathcal{Q}-\pi_{ii}Q_{i}\Big]\xi(s)ds} & +\; \displaystyle{\pi_{ii}\int_{t-h_{1}}^{t}\xi^{\intercal}(s)Q_{i}\xi(s)ds} \\
																										  & \leq \displaystyle{\int_{t-h_{M}}^{t}\xi^{\intercal}(s)\mathcal{Q}\xi(s)ds} & +\; \displaystyle{\eta\int_{t-h_{M}}^{t-h_{1}}\xi^{\intercal}(s)Q_{\kappa}\xi(s)ds} 
\end{array} \label{eq:UppBound1}
\end{align}
where $\sum_{j=1}^{M}\pi_{ij}Q_{j}\leq\mathcal{Q}$. A similar upper bound is readily established for $\sum_{j=1}^{M}\pi_{ij}\int_{t-h_{j+1}}^{t}\dot{\xi}^{\intercal}(s)R\dot{\xi}(s)ds$. We also bound
\begin{small}
\begin{align}
\renewcommand{\arraystretch}{2}
\setlength{\arraycolsep}{1.5pt}
\begin{array}{rll}
\displaystyle{\sum_{j=1}^{M}\pi_{ij}h_{j} \int_{-h_{j}}^{0} \int_{t+s}^{t} \dot{\xi}^{\intercal}(\theta) S\dot{\xi}(\theta)d\theta ds} & =\displaystyle{\sum_{j\neq i}^{M}\pi_{ij}h_{j}\int_{-h_{j}}^{0} \int_{t+s}^{t} \dot{\xi}^{\intercal}(\theta) S \dot{\xi}(\theta)d\theta ds} & +\;\displaystyle{\pi_{ii}h_{i}\int_{-h_{i}}^{0} \int_{t+s}^{t} \dot{\xi}^{\intercal}(\theta) S \dot{\xi}(\theta)d\theta ds} \\
 & \leq \displaystyle{-\pi_{ii}h_{M}\int_{-h_{M}}^{0} \int_{t+s}^{t} \dot{\xi}^{\intercal}(\theta) S \dot{\xi}(\theta)d\theta ds} & +\;\displaystyle{\pi_{ii}h_{1}\int_{-h_{1}}^{0} \int_{t+s}^{t} \dot{\xi}^{\intercal}(\theta) S \dot{\xi}(\theta)d\theta ds} \\ 
 & = \displaystyle{-\pi_{ii}h_{M}\int_{-h_{M}}^{-h_{1}} \int_{t+s}^{t} \dot{\xi}^{\intercal}(\theta) S \dot{\xi}(\theta)d\theta ds} & -\;\displaystyle{\pi_{ii}\big(h_{M}-h_{1}\big)\int_{-h_{1}}^{0} \int_{t+s}^{t} \dot{\xi}^{\intercal}(\theta) S \dot{\xi}(\theta)d\theta ds} \\ 
 & \leq\displaystyle{\eta h_{M}\int_{-h_{M}}^{-h_{1}} \int_{t+s}^{t} \dot{\xi}^{\intercal}(\theta) S \dot{\xi}(\theta)d\theta ds} &+\;\displaystyle{\eta\big(h_{M}-h_{1}\big)\int_{-h_{1}}^{0} \int_{t+s}^{t} \dot{\xi}^{\intercal}(\theta) S \dot{\xi}(\theta)d\theta ds}
\end{array} \label{eq:UppBound2}
\end{align}
\end{small}
and
\begin{align}
\sum_{j=1}^{M}\pi_{ij}(h_{j+1}-h_{j})\int_{-h_{j+1}}^{-h_{j}} \int_{t+s}^{t} \dot{x}^{\intercal}(\theta) T \dot{x}(\theta)d\theta ds \leq &\; \eta\delta_{\max}\int_{-h_{M+1}}^{-h_{1}}\int_{t+s}^{t} \dot{x}^{\intercal}(\theta) T \dot{x}(\theta)d\theta ds \;.
\label{eq:UppBound3}
\end{align}
Now, substituting~\eqref{eq:Jensen2}~--~\eqref{eq:UppBound3} into~\eqref{eq:Vdot1}~--~\eqref{eq:Vdot5}, we get the following inequality
\begin{align}
\mathcal{A}V(\xi_{t},i) \leq &\; \xi^{\intercal}(t)\Bigg[ A_{\alpha}^{\intercal}P_{i}+P_{i}A_{\alpha}+\sum_{j=1}^{2}\pi_{ij}P_{j} + \big(Q_{i} + h_{2}\mathcal{Q} + \delta_{1}Q_{\kappa}\big) + \big(R_{i} + h_{3}\mathcal{R} + \delta_{2}R_{\kappa}\big) - S + A_{\alpha}^{\intercal}\big(\epsilon_{1,i}S+\epsilon_{2,i}T\big)A_{\alpha} \Bigg]\xi(t) \nonumber\\
&\; + 2\xi^{\intercal}(t)\Bigg[ P_{i} + A_{\alpha}^{\intercal}\big(\epsilon_{1,i}S+\epsilon_{2,i}T\big)\Bigg]\sum_{j=1}^{2}(\lambda_{j}(t)A_{\alpha_{ij}})\xi\big(t-\tau_{i}(t)\big) - \xi^{\intercal}(t-h_{i})\big(Q_{i}+S+T\big)\xi(t-h_{i}) +2\xi^{\intercal}(t)S\xi\big(t-h_{i}\big) \nonumber\\ 
&\; - \xi^{\intercal}(t-h_{i+1})\big(R_{i}+T\big)\xi(t-h_{i+1}) +\xi^{\intercal}\big(t-\tau_{i}(t)\big)\big(-2T+Z+Z^{\intercal}\big)\xi\big(t-\tau_{i}(t)\big) + 2\xi^{\intercal}\big(t-\tau_{i}(t)\big)\big(T-Z^{\intercal}\big)\xi\big(t-h_{i}\big)  \nonumber\\
&\; + 2\xi^{\intercal}\big(t-\tau_{i}(t)\big)\big(T-Z\big)\xi\big(t-h_{i+1}\big) +2\xi^{\intercal}\big(t-h_{i}\big)Z\xi\big(t-h_{i+1}\big)\nonumber\\
&\; + \xi^{\intercal}(t-\tau_{i}(t))\sum_{j=1}^{2}(\lambda_{j}(t)A_{\alpha_{ij}})^{\intercal}\big(\epsilon_{1,i}S+\epsilon_{2,i}T\big)\sum_{j=1}^{2}(\lambda_{j}(t)A_{\alpha_{ij}})x(t-\tau_{i}(t)) \nonumber\\ 
= &\; \psi^{\intercal}(t)\tilde{\Gamma}_{i}(t)\psi(t) \;,
\end{align}
where $\psi(t)=\mbox{col}\big\{ \xi(t),\xi(t-\tau_{i}(t)),\xi(t-h_{i}),\xi(t-h_{i+1}) \big\}$. 
Note that $\mathcal{A}V(\xi_{t},r_{t})$ is bounded by a quadratic function in $\psi(t)$:
\begin{equation}
\mathcal{A}V(\xi_{t},i) \leq \psi^{\intercal}(t)\widetilde{\Gamma}_{i}\psi(t) \;,
\end{equation}
where
\begin{equation}
\widetilde{\Gamma}_{ij} = \lambda_{1}(t)\widetilde{\Gamma}_{i1} + \lambda_{2}(t)\widetilde{\Gamma}_{i2}
\end{equation}
for all $i\in\{1,2\}$ and
\begin{equation}
\renewcommand{\arraystretch}{1.2}
\setlength{\arraycolsep}{1.25pt}
\widetilde{\Gamma}_{ij} =
\left[\begin{array}{c:c:cc}
\Phi_{i} & P_{i}A_{\alpha_{ij}} & S & 0 \\ \hdashline
\star & -2T+Z+Z^{\intercal} & T-Z^{\intercal} & T-Z \\ \hdashline
\star & \star & -(Q_{i}+S+T) & Z \\ 
\star & \star & \star & -(R_{i}+T)
\end{array}\right] + \phi_{ij}^{\intercal}\Big( \epsilon_{3,i}S + \epsilon_{4,i}T \Big)\phi_{ij} \;,
\end{equation}
where $\phi_{ij}\triangleq\big[~A~A_{\alpha_{ij}}~0_{n\times 2n}~\big]$ for all $i,j\in\{1,2\}$.
As in Section~2, applying the Schur complement lemma to $\widetilde{\Gamma}_{ij}$ to form $\Gamma_{ij}$, we arrive at the equivalent condition 
\begin{equation}
\Gamma_{i}(t)=\lambda_{1}(t)\Gamma_{i1} + \lambda_{2}(t)\Gamma_{i2}< 0\;, \quad \forall i\in\{1,2\}\,.
\end{equation}
This condition is satisfied for all $\lambda_{i}$ if $\Gamma_{i1}$ and $\Gamma_{i2}$ are both negative definite. Hence, satisfying the condition $\Gamma_{ij}<0$, we guarantee  that the dynamics is exponentially stable with decay rate $\alpha$. This concludes the proof.\hfill $\square$

\begin{remark}
In order to investigate the stability of the Markovian jump system~\eqref{eq:MJLS} with $M$ modes, we must examine $2M$ LMIs (\emph{e.g.},~\eqref{eq:MJLS_Stability_LMI}) whose dimensions are $6n\times 6n$ and, additionally, $2M+1$ small size LMIs (\emph{e.g.},~\eqref{eq:MJLS_Stability_add_LMI}). In total, we use $3M+5$ matrix variables, each with $n(n+1)/2$ decision variables.
\end{remark}

\subsection{State-Feedback Controller Design} \label{subsec:MJSControllerDesign}
Hereafter we concentrate our interest on extending analysis conditions to state-feedback synthesis for exponential mean-square stability of the Markovian jump linear system scheme introduced in Section~\ref{subsec:StochSystemModel}. To this end, we consider a linear time-invariant plant 
\begin{equation}
\dot{x}(t) = Ax(t) + Bu(t) \label{eq:Closed_Loop_System}
\end{equation}
where $x(t)\in\mathbb{R}^{n}$ is the state and $u(t)\in\mathbb{R}^{m}$ is the control input being mode-dependent linear feedback of the delayed state with the following control law
\begin{equation}
u(t) = K_{i}x(t-\tau_{i}(t)) \label{eq:StochControl}
\end{equation}
when $r(t) = i$ (and hence, $\tau_{i}(t)\in[h_{i},h_{i+1})$), $i\in\mathcal{M}$. Indeed, the design problem is to determine a set of state-feedback gain matrices $K_{i}$ that guarantees the closed-loop control stability in exponentially mean-squared sense for the transition rates $\Pi = [\pi_{ij}]_{i,j=1,\cdots,M}$. This problem is closely related to the stability analysis problem discussed in Section~\ref{subsec:StocStability} because the control system structure can also be represented as a Markovian jump linear system on the form of~\eqref{eq:MJLS} with $A_{r(t)}=BK_{r(t)}$. We have the following result.

\begin{thm}\label{thm:MJLS_Stabilization}
For a given decay rate $\alpha >0$, there exists a state-feedback control of the form~\eqref{eq:StochControl} that stabilizes system~\eqref{eq:MJLS} for randomly varying delays $\tau_{i}\in[h_{i},h_{i+1}),~\forall i\in\{1,2\}$ in exponentially mean-squared sense if there exist real constant matrices $\tilde{P}_{i},\tilde{Q}_{i},\tilde{R}_{i}\in\mathbb{S}_{++}^{n},~\forall i\in\{1,2\}$, $\tilde{S},\tilde{T},\tilde{\mathcal{Q}},\tilde{\mathcal{R}}\in\mathbb{S}_{++}^{n}$ and $\tilde{X},\tilde{Z}\in\mathbb{R}^{n\times n}$ such that the following LMIs hold for $i,j\in\{1,2\}$
\begin{align}
\begin{bmatrix}\tilde{T} & \tilde{Z} \\ \tilde{Z}^{\intercal} & \tilde{T}\end{bmatrix} \geq 0\;, \qquad \sum_{j=1}^{2}\pi_{ij}\tilde{Q}_{j}\leq\tilde{\mathcal{Q}}\;, \qquad \sum_{j=1}^{2}\pi_{ij}\tilde{R}_{j}\leq\tilde{\mathcal{R}}\;, \label{eq:MJLS_Stabilization_add_LMI}
\end{align}
and

\begin{align}
\renewcommand{\arraystretch}{1.2}
\setlength{\arraycolsep}{1.25pt}
\left[\begin{array}{c:c:c:cc:c:cc}
-\tilde{X}^{\mathcal{S}} & A_{\alpha}\tilde{X}+\tilde{P}_{i} & \varrho_{ij}B\tilde{Y}_{i} & 0 & 0 & \tilde{X} & \sqrt{\epsilon_{1,i}}\tilde{S} & \sqrt{\epsilon_{2,i}}\tilde{T} \\ \hdashline
\star & \tilde{\beth}_{i} & 0 & \tilde{S} & 0 & 0 & 0 & 0 \\ \hdashline
\star & \star & -\tilde{\Upsilon}^{\mathcal{S}} & \tilde{\Upsilon}^{\intercal} & \tilde{\Upsilon} & 0 & 0 & 0  \\ \hdashline
\star & \star & \star & -(\tilde{Q}_{i}+\tilde{S}+\tilde{T}) & \tilde{Z} & 0 & 0 & 0 \\ 
\star & \star & \star & \star & -(\tilde{R}_{i}+\tilde{T}) & 0 & 0 & 0 \\ \hdashline
\star & \star & \star & \star & \star & -\tilde{P}_{i} & -\sqrt{\epsilon_{1,i}}\tilde{S} & -\sqrt{\epsilon_{2,i}}\tilde{T} \\ \hdashline
\star & \star & \star & \star & \star & \star & -\tilde{S} & 0 \\
\star & \star & \star & \star & \star & \star & \star & -\tilde{T}
\end{array}\right] < 0 \;,
\label{eq:MJLS_Stabilization_LMI}
\end{align}
where $\tilde{\beth}_{i}=\sum_{j=1}^{2} \pi_{ij}\tilde{P}_{j}+\big(\tilde{Q}_{i} + h_{2}\tilde{\mathcal{Q}} + \delta_{1}\tilde{Q}_{\kappa}\big) + \big(\tilde{R}_{i} + h_{3}\tilde{\mathcal{R}} + \delta_{2}\tilde{R}_{\kappa}\big)-\tilde{P}_{i}-\tilde{S}$, $\epsilon_{1,i}\triangleq h_{i}^{2} + \eta\frac{h_{2}^{3}-h_{1}^{3}}{2}$ and $\epsilon_{2,i}\triangleq \delta_{i}^{2} + \eta\delta_{\max}\frac{h_{3}^{2}-h_{1}^{2}}{2}$ with $\eta\triangleq \max\vert\pi_{ii}\vert$, $\kappa\triangleq\mathrm{argmax}\vert\pi_{ii}\vert$ and $\delta_{\max}=\max\vert h_{i+1}-h_{i} \vert,~\forall i\in\{1,2\}$, and $\tilde{\Upsilon}=\tilde{T}-\tilde{Z}$. A stabilizing control law is given by~\eqref{eq:StochControl} with gain $K_{i}=\tilde{Y}_{i}\tilde{X}^{-1}$ for all $i\in\{1,2\}$.
\end{thm}

\noindent\textbf{Proof.} The proof of Theorem~\ref{thm:MJLS_Stabilization} is similar to that of Theorem~\ref{thm:Switched_System_Stabilization}  but an outline of the proof is included for completeness. Again, the structure of~\eqref{eq:MJLS_Stability_LMI} is not adapted to the controller design due to the existance of the multiple product terms $A_{\alpha}S$, $A_{\alpha}T$, $A_{\alpha_{ij}}S$ and $A_{\alpha_{ij}}T$ preventing to find a linearizing change of variable even after congruence transformations. As a result, a relaxation approach is applied (as in \S~\ref{sec:SwitchedSyssSynthesis}) to remove the multiple product terms preventing the change of variables. We let~\eqref{eq:MJLS_Stabilization_LMI} be called $\Psi_{ij}$ and we prove the condition $\Psi_{ij}<0$. Similarly to $\Theta_{ij}$ (but with the difference that we have $X$ instead of $X_{i}$), $\Psi_{ij}$ can be decomposed as follows: 
\begin{align*}
\Psi_{ij}=\Psi_{ij}\vert_{X=0}+U_{i}^{\intercal}XV_{i}+V_{i}^{\intercal}X^{\intercal}U_{i}< 0, \quad \forall i\in\{1,2\}
\end{align*}
where $U_{i}=\big[-I_{n}~A_{\alpha}~A_{\alpha_{ij}}~0_{n\times 2n}~I_{n}~0_{n\times 2n}\big]$ and $V_{i}=\big[I_{n}~0_{n\times 7n}\big]$. Then invoking the projection lemma~\cite{GaA:94}, the feasibility of $\Psi_{ij}< 0$ implies the feasibility of the LMIs
\begin{align*}
\mathcal{N}_{U_{i}}^{T}\Psi_{ij}|_{X=0}\mathcal{N}_{U_{i}}< 0 \;,  \\
\mathcal{N}_{V_{i}}^{T}\Psi_{ij}|_{X=0}\mathcal{N}_{V_{i}}< 0 \;,
\end{align*}
where $\mathcal{N}_{U_{i}}$ and $\mathcal{N}_{V_{i}}$ are basis of the null space of $U_{i}$ and $V_{i}$, respectively. Subsequently, the inequality~\eqref{eq:MJLS_Application_of_Projection_Lemma_LMI} is obtained as
\begin{equation}
\renewcommand{\arraystretch}{1.2}
\setlength{\arraycolsep}{1.25pt}
\left[\begin{array}{c:c:c:cc:c:cc}
-X^{\mathcal{S}} & X^{\intercal}A_{\alpha}+P_{i} &X^{\intercal}A_{\alpha_{ij}} & 0 & 0 & X^{\intercal} & \sqrt{\epsilon_{1,i}}S & \sqrt{\epsilon_{2,i}}T \\ \hdashline
\star & \beth_{i} & 0 & S & 0 & 0 & 0 & 0 \\ \hdashline
\star & \star & -\Upsilon^{\mathcal{S}} & \Upsilon^{\intercal} & \Upsilon & 0 & 0 & 0  \\ \hdashline
\star & \star & \star & -(Q_{i}+S+T) & Z & 0 & 0 & 0 \\ 
\star & \star & \star & \star & -(R_{i}+T) & 0 & 0 & 0 \\ \hdashline
\star & \star & \star & \star & \star & -P_{i} & -\sqrt{\epsilon_{1,i}}S & -\sqrt{\epsilon_{2,i}}T\\ \hdashline
\star & \star & \star & \star & \star & \star & -S & 0 \\
\star & \star & \star & \star & \star & \star & \star & -T
\end{array}\right] < 0 \;.
\label{eq:MJLS_Application_of_Projection_Lemma_LMI}
\end{equation}

We substitute the closed-loop system~\eqref{eq:Closed_Loop_System} into the inequality~\eqref{eq:MJLS_Stability_LMI}, and introduce a constant matrix $X\in\mathbb{R}^{n\times n}$. Then, we compel $X$ to be constant, and we perform a congruence transformation with respect to matrix $I_{8n}\otimes X^{-1}$ and apply the following linearizing change of variables $\tilde{X}\triangleq X^{-1},~\tilde{P}_{i}\triangleq \tilde{X}^{\intercal}P_{i}\tilde{X},~\tilde{Q}_{i}\triangleq \tilde{X}^{\intercal}Q_{i}\tilde{X},~\tilde{\mathcal{Q}}\triangleq \tilde{X}^{\intercal}\mathcal{Q}\tilde{X},~\tilde{R}_{i}\triangleq \tilde{X}^{\intercal}R_{i}\tilde{X},~\tilde{\mathcal{R}}\triangleq \tilde{X}^{\intercal}\mathcal{R}\tilde{X},~\tilde{S}\triangleq \tilde{X}^{\intercal}S\tilde{X},~\tilde{T}\triangleq \tilde{X}^{\intercal}T\tilde{X},~\tilde{\Xi}_{i1}\triangleq\tilde{X}^{\intercal}\Xi_{i1}\tilde{X},~\tilde{\Xi}_{i2}\triangleq\tilde{X}^{\intercal} \Xi_{i2}\tilde{X},~\tilde{\Xi}_{i3}\triangleq \tilde{X}^{\intercal}\Xi_{i3}\tilde{X},~\tilde{Z}\triangleq \tilde{X}^{\intercal}Z\tilde{X}$ and $\tilde{Y}_{i}=K_{i}\tilde{X},~\forall i\in\{1,2\}$ in~\eqref{eq:MJLS_Application_of_Projection_Lemma_LMI}, LMI~\eqref{eq:MJLS_Stabilization_LMI} is derived. \hfill $\square$

\begin{remark}
To design a set of stabilizing controllers for Markovian jump system (as discussed in \S~\ref{subsec:MJSControllerDesign}) with $M$ modes, $2M$ LMIs (\emph{e.g.},~\eqref{eq:MJLS_Stabilization_LMI}) whose dimensions are $8n\times 8n$, and also $2M+1$ small size LMIs (\emph{e.g.},~\eqref{eq:MJLS_Stabilization_add_LMI}) must be checked. All these LMIs include $3M+6$ matrix variables with $n(n+1)/2$ decision variables.
\end{remark}

\section{Numerical Examples}\label{sec:design_example}

We are now ready to demonstrate the proposed technique on numerical examples. Our first example, a simple DC-motor model, is included to demonstrate the flexibility of our control structure  compared to a single robust controller. The second example is taken from wide-area power systems, and demonstrates that the numerical techniques scale to non-trivial system dimensions.  Finally, we return to the small-scale example to illustrate the analysis procedures for the random Markovian delay model.
\begin{figure}\centering
  	\includegraphics{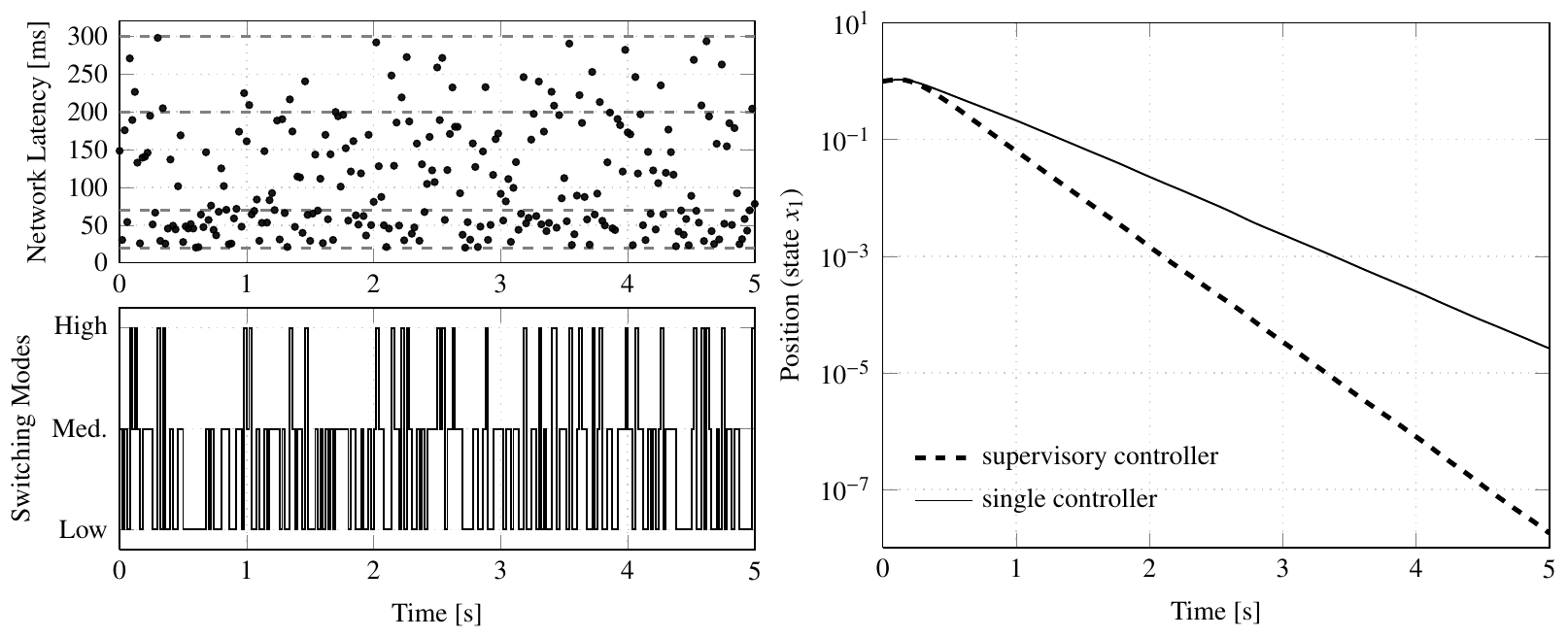}
    \caption{The left figures illustrate a sample evolution of the time delay and the associated delay bounds that define the two controller modes; the associated mode evolution is shown in the bottom left figure. The right figure shows a representative state trajectory of the closed-loop system under supervisory control (dashed line) and mode-independent state-feedback (solid). Both simulations are performed from the same initial value.}
    \label{fig:DC_Motor_Example}
\end{figure} 
\subsection{Small-scale Example: DC Motor}
Consider the following linear system
\begin{align}
\dot{x}(t) = \begin{bmatrix} 0 & 1 \\ 0 & -10 \end{bmatrix}x(t) + \begin{bmatrix} 0 \\ 0.024 \end{bmatrix}u(t) \label{eq:DC_Motor}
\end{align}
with a time-varying communication delay between sensor and controller that behaves as shown in Figure~\ref{fig:DC_Motor_Example}. The supervisor generates the switching signals shown in Figure~\ref{fig:DC_Motor_Example} and triggers the most appropriate controller according to
\begin{align}
u(t) = \begin{cases}
K_{L}x\big(t-\tau_{L}(t)\big) & \mbox{if}~\tau_{L}(t)\in[20,70)\,\mathrm{ ms}\;, \\
K_{M}x\big(t-\tau_{M}(t)\big) & \mbox{if}~\tau_{M}(t)\in[70,200)\,\mathrm{ ms}\;, \\
K_{H}x\big(t-\tau_{H}(t)\big) & \mbox{if}~\tau_{H}(t)\in[200,300)\,\mathrm{ ms}\;.
\end{cases}
\end{align}

By solving the optimization problem~\eqref{eq:optprob} for $\mu=1.4$, we find the lower bound on the average dwell-time $\tau_{a}^{\circ}=0.12\mathrm{s}$ and the corresponding convergence rate $\alpha_{\circ}=2.78$ guaranteed for the gains
\begin{align*}
	K_{L} =&\; \begin{bmatrix} -1421.0 & -138.9 \end{bmatrix}\;, \\
	K_{M} =&\; \begin{bmatrix} -1035.9 & -101.5 \end{bmatrix}\;, \\
	K_{H} =&\; \begin{bmatrix} -757.09 & -72.71 \end{bmatrix}\;.
\end{align*}
Using the same class of Lyapunov-Krasovskii functionals, we design a classical state-feedback controller to compare the non-switching and switching control performance. Specifically, we use the Lyapunov-Krasovskii functional~\eqref{eq:LyapCand} with $Q_{k}=0$, $R_{k}=0$, $S_{k}=0$, $T_{k}=0,~\forall k>1$ and $\tau(t)\in[20,300)$ ms, and find $\alpha=1.72$ for the state feedback gain
\begin{equation*}
K = \begin{bmatrix} -765.74 & -75.74 \end{bmatrix}\;.
\end{equation*}
We observe that the switching controller has a better performance than the non-switching one, $\alpha_{\circ}=2.78>\alpha=1.72$, and that the controller in the low-delay mode can be made much more agressive when we use a mode-dependent controller. The improved convergence rate is confirmed by the simulations shown in Figure~\ref{fig:DC_Motor_Example}.

For different values of $h_{2}$ and $h_{3}$, which define the boundary between the low-, medium- and high-delay mode, we compute the exponential decay rate $\alpha$ of the supervisory controller, and then try to find the largest $h_{\max}$ for which a single mode-independent controller can guarantee the same decay rate. The results are depicted in Table~\ref{tab:table_example}. As can be seen, increasing $h_{2}$ and decreasing $h_{3}$ allow to guarantee an improved decay rate of the switching controller. To guarantee the same decay rate under a mode-indenendent controller, the maximum delay must be reduced, and sometimes significantly so.
%
\begin{table*}[!htbp]
\renewcommand{\arraystretch}{1.2}
\centering
\begin{tabular}
{
 >{\centering \arraybackslash } c <{}
 >{\centering \arraybackslash \columncolor{black!05!white}} c <{}
 >{\centering \arraybackslash \columncolor{black!05!white}} c <{}
 >{\centering \arraybackslash \columncolor{black!05!white}} c <{}
 >{\centering \arraybackslash \columncolor{black!05!white}} c <{}
 >{\centering \arraybackslash } c <{}
 >{\centering \arraybackslash } c <{}
}
 \toprule
   &
 \multicolumn{4}{c}{Supervisory Control} &
 \multicolumn{2}{c}{Non--Switching Control} \\
 \midrule
 \bfseries $\alpha$ &
 \bfseries $h_{1}$ &
 $h_{2}$ &
 $h_{3}$ & 
 $h_{4}$ &
 $h_{\min}$ &
 $h_{\max}$  \\
 \midrule
3.00 &
20 &
100 &
200 &
300 & 
20 &
158 \\
2.42 &
20 &
100 &
250 &
300 & 
20 &
204 \\
2.78 &
20 &
70 &
200&
300 & 
20 &
173 \\
2.27 &
20 &
70 &
250 &
300 &  
20 &
219 \\
 \bottomrule
\end{tabular}
\caption{Comparision switching vs. non-switching controllers for different decay rates and delay intervals} \label{tab:table_example}
\end{table*}

\subsection{Large Scale Example: Wide-Area Power Networks}
%
\begin{figure}\centering
    \includegraphics[angle=0,width=0.4\hsize]{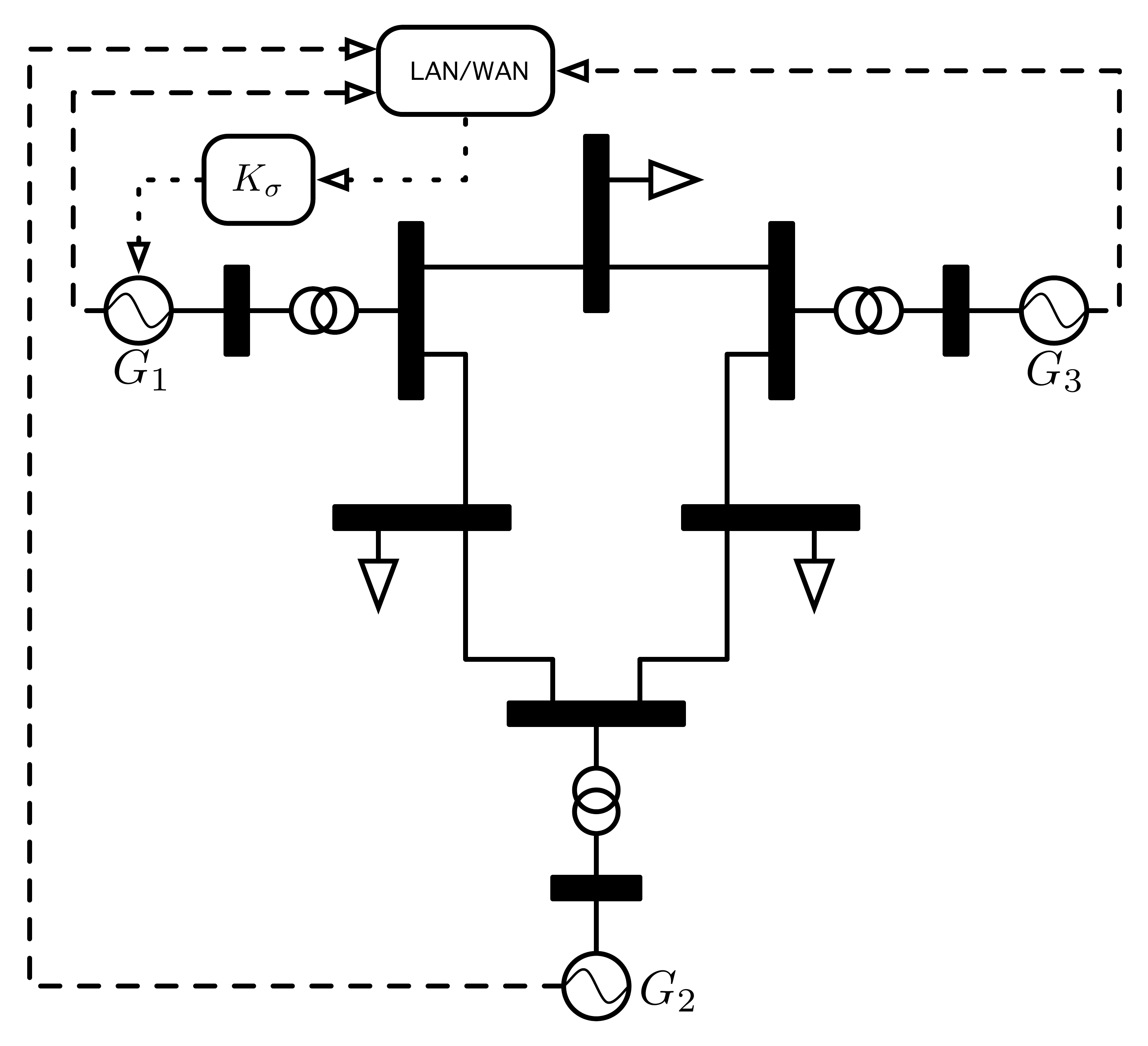}\\
    \caption{IEEE nine-bus power system.}
    \label{fig:ninebus}
\end{figure}
%
To demonstrate the applicability of our methods to systems of higher dimension, we consider the IEEE nine-bus system~\cite{AnF:03} shown in Figure~\ref{fig:ninebus}. We adopt a second order (swing) model with phase and frequency $(\delta_{i}$, $\omega_{i})$ for all generators and use the Power System Analysis Toolbox~\cite{Mil:10} to obtain the following numerical model
\begin{align}
\renewcommand{\arraystretch}{1.2}
\frac{d}{dt}
\left[\begin{array}{c}
\delta_{1} \\ \omega_{1} \\ \delta_{2} \\ \omega_{2} \\ \delta_{3} \\ \omega_{3} 
\end{array}\right] = 
\left[\begin{array}{cccccc}
 0 & 1 & 0 & 0 & 0 & 0 \\
 -0.0432 & -0.0702 & 0.0209 & 0 & 0.0223 & 0 \\
 0 & 0 & 0 & 1 & 0 & 0 \\
 0.1248 & 0 & -0.2372 & -0.2594 & 0.1124 & 0 \\
 0 & 0 & 0 & 0 & 0 & 1 \\
 0.3761 & 0 & 0.3554 & 0 & -0.7315 & -0.5515
\end{array}\right]
\left[\begin{array}{c}
\delta_{1} \\ \omega_{1} \\ \delta_{2} \\ \omega_{2} \\ \delta_{3} \\ \omega_{3} 
\end{array}\right]  +
\left[\begin{array}{c}
0 \\ 0.1471 \\ 0 \\ 0 \\ 0 \\ 0 
\end{array}\right]u(t) \;. \label{eq:ninebus}
\end{align}

\begin{figure}\centering
  	\includegraphics{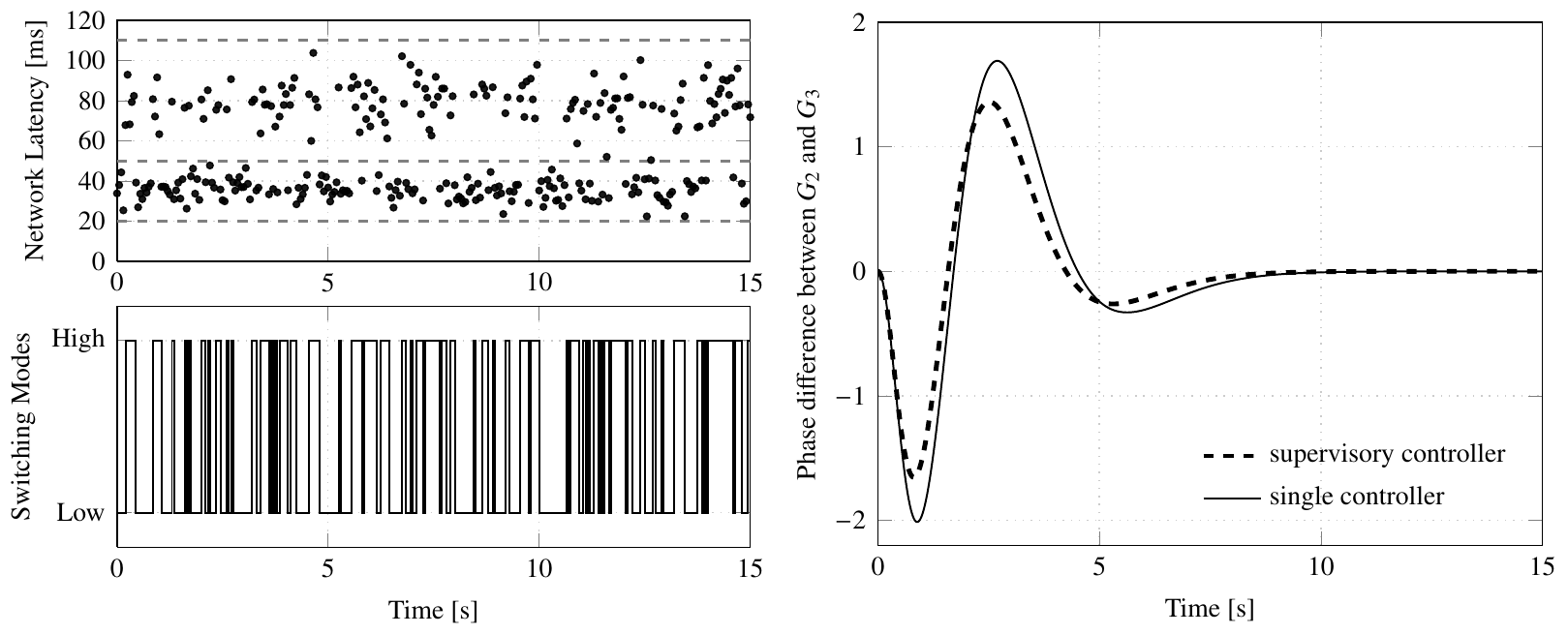}
    \caption{The left figures show a sample delay evolution and the delay bounds that define the supervisory control modes (top) along with the associated mode evolution (bottom). The right figure illustrates a representative state trajectory of the closed-loop system under supervisory control (dashed line) and a single mode-independent state-feedback (solid). Both simulations are performed from the same initial value.}
    \label{fig:Delay_Evolution}
\end{figure}

We assume that the phase and frequency of each bus can be measured and be communicated to a central controller. In wide-area power systems, the communication delays vary depending on communication technologies, protocols and network load. In this example, we assume that the delay varies between 20 and 110 ms (see, Figure~\ref{fig:Delay_Evolution}) and that mode changes are such that the average dwell-time is guaranteed to be at least 0.35 seconds. 

To begin, the supervisory controller is designed for the delay intervals [20, 50) and [50,110) ms. We design supervisory control gains targeting a decay rate of $\alpha=0.9$, and use $\mu=1.4$ to guarantee $\tau_a=0.3739$. As a result, we compute the following control gain matrices
\begin{align*}
K_{L} = \begin{bmatrix} -181.60 & -53.94 & -83.97 & -2090.02 & -647.79 & -85.90 \end{bmatrix}\;, \\
K_{H} = \begin{bmatrix} -159.82 & -49.22 & -48.09 & -1714.38 & -547.06 & -94.92 \end{bmatrix}\;.
\end{align*}
Solving the LMIs for this larger system takes a few hours, but appears to be numerically stable. 

It turns out that an exponential decay rate of $\alpha = 0.9$ can also be achieved by the mode-independent feedback gain
\begin{equation}
K = 
\begin{bmatrix}
  -153.71 & -48.22 & -41.69 & -1605.33 & -514.67 & -96.07
\end{bmatrix} \;,
\end{equation}
which is less aggressive than the controller used in the low-delay mode of the switching controller. As depicted in Figure~\ref{fig:Delay_Evolution}, simulations are carried out for the delay trace and corresponding switching singnals shown in the left of the same figure, the switching controller damps oscillations between generators $G_2$ and $G_3$ better than its mode-independent counterpart.

\subsection{Markovian Jump Linear System Formulation}
Finally, we return to the DC-motor example to illustrate the applicability of the analysis tools developed in~Section~\ref{sec:StocSwitchedSys} for random delays. We consider the system~\eqref{eq:DC_Motor} with low traffic delays (\emph{i.e.}, $\tau_{L}\in[20,70)$ms) and high traffic delays (\emph{i.e.}, $\tau_{H}\in[70,300)$ms). The switching between these two modes is described by the following transition probability rate matrix: 
\begin{equation*}
\Pi = 
\begin{bmatrix}
-p & p \\ q & -q
\end{bmatrix} \;.
\end{equation*}
whose invariant distribution is $\pi_{1}^{\infty}=q/(p+q)$ and $\pi_{2}^{\infty}=p/(p+q)$. We vary the transition probabilities $p$ and $q$, compute mode-dependent controllers and the rate of exponential mean-square stability. The results are summerized in Table~\ref{tab:Comparision_of_MJLS}. While mitigating $\max\big\{p,q\big\}$, the decay rate is increasing and controllers are getting more aggressive. Furthermore, when $\pi_{1}^{\infty}$ decreases, we also observe a slight decrease in the decay rate.
%
\begin{table*}[!htbp]
\renewcommand{\arraystretch}{1.2}
\centering
\begin{tabular}
{
 >{\centering \arraybackslash } c <{}
 >{\centering \arraybackslash \columncolor{black!05!white}} c <{}
 >{\centering \arraybackslash \columncolor{black!05!white}} c <{}
 >{\centering \arraybackslash } c <{}
 >{\centering \arraybackslash } c <{}
 >{\centering \arraybackslash \columncolor{black!05!white}} c <{}
 >{\centering \arraybackslash \columncolor{black!05!white}} c <{}
}
 \toprule
   &
 \multicolumn{2}{c}{Invariant} &
 \multicolumn{2}{c}{Transition} &
 \multicolumn{2}{c}{Mode-dependent} \\
   &
 \multicolumn{2}{c}{Distributions} &
 \multicolumn{2}{c}{Probabilities} &
 \multicolumn{2}{c}{Controllers} \\
 \midrule
 \bfseries $\alpha$ &
 \bfseries $\pi_{1}^{\infty}$ &
 $\pi_{2}^{\infty}$ &
 $p$ & 
 $q$ &
 $K_{L}$ &
 $K_{H}$  \\
 \midrule
 $1.07$ & 
 $12.50 $ & 
 $87.50$ & 
 $3.5$ & 
 $0.5$ & 
 $\big[-651.93~-63.80\big]$ & 
 $\big[-542.15~-53.04\big]$ \\
 $1.08$ & 
 $36.36 $ & 
 $63.64$ & 
 $3.5$ & 
 $2.0$ & 
 $\big[-652.71~-63.87\big]$ & 
 $\big[-537.72~-52.60\big]$\\
 $1.09$ & 
 $50.00 $ & 
 $50.00$ & 
 $3.5$ & 
 $3.5$ & 
 $\big[-654.28~-64.50\big]$ & 
 $\big[-536.68~-52.89\big]$\\
 $1.10$ & 
 $63.64 $ &
 $36.36$ & 
 $2.0$ & 
 $3.5$ & 
 $\big[-652.41~-65.18\big]$ & 
 $\big[-539.49~-53.90\big]$\\
 $1.10$ & 
 $87.50 $ & 
 $12.50$ & 
 $0.5$ & 
 $3.5$ & 
 $\big[-656.22~-64.04\big]$ & 
 $\big[-503.26~-49.07\big]$\\ 
 \midrule
 $1.23$ & 
 $16.67$ & 
 $83.33$ & 
 $2.5$ & 
 $0.5$ & 
 $\big[-725.08~-72.40\big]$ & 
 $\big[-592.00~-59.11\big]$\\ 
 $1.25$ & 
 $37.50$ & 
 $62.50$ & 
 $2.5$ & 
 $1.5$ & 
 $\big[-726.46~-72.12\big]$ & 
 $\big[-586.40~-58.21\big]$\\
 $1.25$ & 
 $50.00$ & 
 $50.00$ & 
 $2.5$ & 
 $2.5$ & 
 $\big[-728.26~-72.33\big]$ & 
 $\big[-583.34~-57.93\big]$\\
 $1.25$ &
 $62.50$ & 
 $37.50$ & 
 $1.5$ & 
 $2.5$ & 
 $\big[-721.48~-71.11\big]$ & 
 $\big[-579.53~-57.09\big]$\\
 $1.26$ &
 $83.33$ &
 $16.67$ &
 $0.5$ & 
 $2.5$ & 
 $\big[-715.62~-71.20\big]$ & 
 $\big[-581.17~-57.81\big]$\\
 \midrule
 $1.43$ & 
 $25.00$ & 
 $75.00$ & 
 $1.5$ & 
 $0.5$ & 
 $\big[-804.16~-80.05\big]$ & 
 $\big[-645.74~-64.27\big]$\\
 $1.46$ &
 $75.00$ &
 $25.00$ &
 $0.5$ &
 $1.5$ &
 $\big[-798.94~-79.58\big]$ &
 $\big[-639.16~-63.66\big]$\\
 $1.45$ & 
 $50.00$ & 
 $50.00$ & 
 $1.5$ & 
 $1.5$ & 
 $\big[-810.53~-80.11\big]$ & 
 $\big[-637.13~-62.95\big]$\\
 \bottomrule
\end{tabular}
\caption{Results of Markovian jump linear controller design for $h_{1}=20\mathrm{ms}$, $h_{2}=70\mathrm{ms}$ and $h_{3}=300\mathrm{ms}$} \label{tab:Comparision_of_MJLS}
\end{table*}

\section{Conclusion}\label{sec:conc}

This paper has been dedicated to supervisory control of networked control systems with time-varying delays. The main contribution of this paper is to develop a  stability analysis and a state feedback synthesis technique for a supervisory control system that switches among a multi-controller unit based on the current network state. A novel class of Lyapunov-Krasovskii functionals were introduced that, somewhat remarkably, admit both the analysis and state feedback synthesis problems to be solved via convex optimization over linear matrix inequalities. In addition, we also investigated the corresponding problem for a class of stochastic systems with interval-bounded time-varying delay. Sufficient conditions were established without ignoring any terms in the weak infinitesimal operator of the Lyapunov Krasovskii functional by considering the relationship among the time-varying delay, its upper bound, and their difference. Finally, examples were given to show the effectiveness of the proposed analysis and synthesis techniques. 


\bibliographystyle{model1-num-names}
\bibliography{ACC12bibtex}

\end{document}